\documentclass[iop]{emulateapj}

\usepackage{my_astro, numbers, graphics, aas_macros}
\usepackage[usenames, dvipsnames]{color}
\usepackage[hidelinks]{hyperref}

\newcommand{\pds}{j}

\newcommand{\drate}{J}
\newcommand{\dratebol}{J}
\newcommand{\cds}{\tilde{j}}
\newcommand{\lamion}{\lambda_\mathrm{ion}}
\newcommand{\pflux}{F_p}
\newcommand{\eflux}{F}
\newcommand{\ebolo}{I}
\newcommand{\xsctn}{\sigma}

\newcommand{\ch}[1]{\colhead{#1}}
\newcommand{\tnm}[1]{\tablenotemark{#1}}
\newcommand{\teff}{$T_\mathrm{eff}$}
\newcommand{\feh}{[Fe/H]}
\newcommand{\logg}{$\log_{10}{g}$}
\newcommand{\hlsplink}{https://archive.stsci.edu/prepds/muscles/} 
\newcommand{\moleculelist}{\HII, \NII, \OII, \OIII, \HIIO, CO, \COII, \CHIV, and \NIIO}

\newcommand{\epseri}{$\mathrm{\epsilon}$ Eri}

\newcommand{\refchng}[2]{#2}
\newcommand{\mychng}[1]{#1}

\begin{document}

\submitted{Version \today, Accepted to The Astrophysical Journal}
\title{The MUSCLES Treasury Survey III: X-ray to Infrared Spectra of 11 M and K Stars Hosting Planets}

\author{R. O. Parke Loyd \altaffilmark{1}, Kevin France\altaffilmark{1}, Allison Youngblood\altaffilmark{1}, Christian Schneider\altaffilmark{2}, Alexander Brown\altaffilmark{3}, Renyu Hu\altaffilmark{4,5}, Jeffrey Linsky\altaffilmark{6}, Cynthia S. Froning\altaffilmark{7}, Seth Redfield\altaffilmark{8}, Sarah Rugheimer\altaffilmark{9}, Feng Tian\altaffilmark{10}}

\altaffiltext{1}{Laboratory for Atmospheric and Space Physics, Boulder, Colorado,  80309; \email{robert.loyd@colorado.edu}}
\altaffiltext{2}{European Space Research and Technology Centre (ESA/ESTEC), Keplerlaan 1, 2201 AZ Noordwijk, The Netherlands}
\altaffiltext{3}{Center for Astrophysics and Space Astronomy, University of Colorado, 389 UCB, Boulder, CO 80309}
\altaffiltext{4}{Jet Propulsion Laboratory, California Institute of Technology, Pasadena, CA 91109}
\altaffiltext{5}{Division of Geological and Planetary Sciences, California Institute of Technology, Pasadena, CA 91125}
\altaffiltext{6}{JILA, University of Colorado and NIST, 440 UCB, Boulder, CO 80309}
\altaffiltext{7}{Dept. of Astronomy C1400, University of Texas, Austin, TX, 78712}
\altaffiltext{8}{Astronomy Department and Van Vleck Observatory, Wesleyan University, Middletown, CT 06459-0123, USA}
\altaffiltext{9}{Dept of Earth and Environmental Sciences, University of St. Andrews, Irvine Building, North Street, St. Andrews KY16 9AL, UK}
\altaffiltext{10}{Ministry of Education Key Laboratory for Earth System Modeling, Center for Earth System Science, Tsinghua University, Beijing, 100084, China}

\begin{abstract}
We present a catalog of panchromatic spectral energy distributions (SEDs) for 7 M and 4 K dwarf stars that span X-ray to infrared wavelengths (5 \AA\ ~--~5.5 \micron). 
These SEDs are composites of \scraft{Chandra} or \scraft{XMM-Newton} data from 5~--~$\sim$50 \AA, a plasma emission model from $\sim$50~--~100 \AA, broadband empirical estimates from 100~---1170 \AA, \scraft{HST} data from 1170~---~5700 \AA, including a reconstruction of stellar \lya\ emission at 1215.67 \AA, and a PHOENIX model spectrum from 5700~--~55000 \AA.
\mychng{
Using these SEDs, we computed the photodissociation rates of several molecules prevalent in planetary atmospheres when exposed to each star's unattenuated flux (``unshielded'' photodissociation rates) and found that rates differ among stars by over an order of magnitude for most molecules. 
In general, the same spectral regions drive unshielded photodissociations both for the minimally and maximally FUV active stars.
However, for \OIII\ visible flux drives dissociation for the M stars whereas NUV flux drives dissociation for the K stars.  
We also searched for an FUV continuum in the assembled SEDs and detected it in \vnContDetections/11 stars, where it contributes around 10\% of the flux in the range spanned by the continuum bands. 
An ultraviolet continuum shape is resolved for the star \epseri\ that shows an edge likely attributable to \Siii\ recombination.
The 11 SEDs presented in this paper, available online through the Mikulski Archive for Space Telescopes, will be valuable for vetting stellar upper-atmosphere emission models and simulating photochemistry in exoplanet atmospheres.}
\end{abstract}

\section{Introduction}

\refchng{1}{Current stellar and planetary population statistics indicate that most rocky planets orbit low-mass stars.} 
Low-mass stars, specifically spectral types M and K, greatly outnumber those of higher-mass, making up at least 91\% of the stellar population within 10 pc of the Sun \citep{henry06}. 
On average, the low-mass stellar population \refchng{1}{exhibits a planetary occurrence rate of} 0.1~--~0.6 habitable zone (HZ) terrestrial planets per star \citep{dressing13, kopparapu13, dressing15}.
\refchng{1}{Further, the occurrence rate of rocky planets was found to decrease with increasing stellar mass by \cite{howard12} and no trend with stellar mass was found by \cite{fressin13}.
The above results collectively imply that,} by numbers alone, low-mass stars are certain to be a cornerstone of exoplanet science and the search for other Earths.

The abundance of low-mass stars ensures that many are close enough to enable high-precision photometry and spectroscopy. 
In addition, their \refchng{2}{smaller sizes and lower masses} yield deeper planetary transits and larger stellar \mychng{reflex radial velocities} when compared to a system with a higher stellar mass but identical planetary mass and orbital period. 
Furthermore, orbits around cooler, less-luminous stars must \mychng{have shorter periods than those around} Sun-like stars to achieve the same planetary effective temperature, making transits more likely and frequent and enhancing reflex velocities for radial velocity detection.
These advantages facilitate the detection and bulk characterization (e.g. mass, radius) of planets orbiting nearby low-mass stars using radial velocity and transit techniques.
They also facilitate atmospheric characterization through transmission spectroscopy, as has recently been performed for super-Earths orbiting the \refchng{3}{M4.5 star GJ 1214 and K1 star HD 97658} \citep{kreidberg14, knutson14}.

Upcoming exoplanet searches, such as the \scraft{Transiting Exoplanet Survey Satellite (TESS)}, will focus on low-mass stars. 
In searching for planets as cool as Earth, \scraft{TESS} is biased by its short (1~--~12 month) monitoring of host stars \citep{ricker14}.
Only around low-mass stars will cool planets \mychng{orbit with short enough periods} to transit several times during these monitoring programs.
As a result, the \scraft{TESS} sample of potentially-habitable exoplanets  will mostly be orbiting low-mass stars \citep{deming09}.

The prevalence of low-mass stars hosting planets makes a thorough knowledge of the typical planetary environment \mychng{provided by such stars} indispensable. 
However, the circumstellar environment of an M or K dwarf differs substantially from the well-studied environment of the Sun. 
Lower-mass stars have cooler photospheres that emit \mychng{spectra} peaking at redder wavelengths in comparison with Sun-like stars, but there are other important differences to address.
As we discuss these differences and throughout this paper, we will refer to various ranges of the stellar spectral energy distributions using their established monikers. 
We adopt the definitions of X-ray $<\ $ \xray , EUV = [\euvlo , \euvhi] (\euvhi\ corresponds to the red edge of emission from H II recombination), XUV (X-ray + EUV) $<\ $ \xuv , FUV = [\fuvlo , \fuvhi], NUV = [\nuvlo , \nuvhi], visible = [\vislo , \vishi], IR $>\ $\ir. 
Although emission from the stellar photosphere dominates the energy budget of emitted stellar radiation ($F(<1700~\mathrm{\AA})/F(>1700~\mathrm{\AA}) \approx 10^{\nonPhotFluxRatio}$ for the stars in this paper), X-ray through UV emission that is emitted primarily from the stellar chromosphere and corona drives photochemistry, ionization, and mass loss in the atmospheres of orbiting planets. 
\refchng{4}{We briefly describe thee influences below.}

FUV photons are able to dissociate molecules, both heating the atmosphere and directly modifying its composition.
This affects many molecules common in planetary atmospheres, including \OII, \HIIO, \COII, and \CHIV. 
If the fraction of FUV to bolometric flux is sufficiently large, photodissociation can push the atmosphere significantly out of thermochemical equilibrium (see, e.g., \citealt{moses14, miguel14, hu14}).
Photochemical reactions can produce a buildup of \OII\ and \OIII\ at concentrations that, in the absence of radiative forcing, might be considered indicators of biological activity \citep{tian14, domagal14, wordsworth14, harman15}. 
It is also possible that these atmospheric effects, besides just interfering with the detection (or exclusion) of existing life, might also influence the emergence and evolution of life. 
UV radiation can both damage \citep{voet63, matsunaga91,tevini93, kerwin07} and aid in synthesizing \citep{senanayake06, barks10, ritson12, patel15} many molecules critical to the function of Earth's life.

At higher photon energies, stellar extreme ultraviolet (EUV) and X-ray photons (often termed XUV in conjunction \mychng{with the EUV}) can eject electrons from atoms, ionizing and heating the upper atmospheres of planets. 
This heating can drive significant atmospheric escape for close-in planets \citep{lammer03, yelle04, tian05, murray09}.
In addition, the development of an ionosphere through ionization by EUV flux will influence the interaction (and associated atmospheric escape) of the planetary magnetosphere with the stellar wind (e.g. \citealt{cohen14}).
Indeed, atmospheric escape has been observed on the hot Jupiters HD 209458b \citep{vidal03, linsky10}, HD 189733b \citep{lecavelier10}, and WASP-12b \citep{fossati10, fossati13} and the hot Neptune GJ 436b \citep{kulow14,ehrenreich15}.

The effects of stellar radiation on planetary atmospheres warrant the spectroscopic characterization of low-mass stars at short wavelengths. 
Yet these observations are rare compared to visible-IR observations and model-synthesized spectra.  
To address this scarcity, \cite{france13} conducted a pilot program collecting UV spectra of 6 M dwarfs, a project termed Measurements of the Ultraviolet Spectral Characteristics of Low-Mass Exoplanetary Systems (MUSCLES).

Responding to the success of the pilot survey and continued urging of the community (e.g. \citealt{segura05, domagal14, tian14, cowan15, rugheimer15}), we have completed the MUSCLES Treasury Survey, doubling the stellar sample and expanding the spectral coverage from $\Delta\lambda \approx 2000$~\AA\ in the ultraviolet to a span of four orders of magnitude in wavelength (5~\AA~--~5.5~\micron).
We have combined  observations in the X-ray, UV, and blue-visible; reconstructions of the ISM-absorbed \lya\ and EUV flux; and spectral output of the PHOENIX photospheric models (\citealt{husser13}; 5700~\AA~--~5.5~\micron) and APEC coronal models (\citealt{smith01}; $\sim$50~--~100~\AA) to create panchromatic spectra for all targets. 
The fully reduced and co-added spectral catalog is publicly available through \href{\hlsplink}{the Mikulski Archive for Space Telescopes (MAST)}.\footnote{\hlsplink}

We present the initial results of the MUSCLES Treasury Survey in three parts:
\citeauthor{france16} (\citeyear{france16}; hereafter ``Paper I'') gives the overview of this MUSCLES Treasury program including an investigation of the dependence of total UV and X-ray luminosities and individual line fluxes on stellar and planetary parameters that reveals a tantalizing suggestion of star-planet-interactions.
\citeauthor{youngblood16} (\citeyear{youngblood16}; hereafter ``Paper II'') supplies the details of the \lya\ reconstruction and EUV modeling and explores the possibility of estimating EUV and \lya\ fluxes from other, more readily observed emission lines and the correlation of these fluxes with stellar rotation.
This paper, the third in the series, discusses the details of assembling the panchromatic spectral energy distributions (SEDs) that are the primary data product of the MUSCLES Treasury Survey, with a report on the detection of FUV continua in the targets, a computation of ``unshielded'' dissociation rates for important molecules in planetary atmospheres, and a comparison to purely photospheric PHOENIX spectra.

This paper is structured as follows: 
Section \ref{sec:data} describes the data reduction process, separately addressing each of the sources of the spectra that are combined into the composite SEDs, the process for this combination, special cases, and overall data quality. 
Section \ref{sec:discuss} explores the SED catalog and discusses its use. 
There we examine the FUV continuum, make suggestions for estimating the SEDs of non-MUSCLES stars, compute and compare unattenuated photodissociation rates, and illustrate the importance of accounting for emission from stellar upper atmospheres in SEDs. Section \ref{sec:summary} then summarizes the data products and results.

\section{The Data}
\label{sec:data}

\begin{figure*}
\centering
\includegraphics{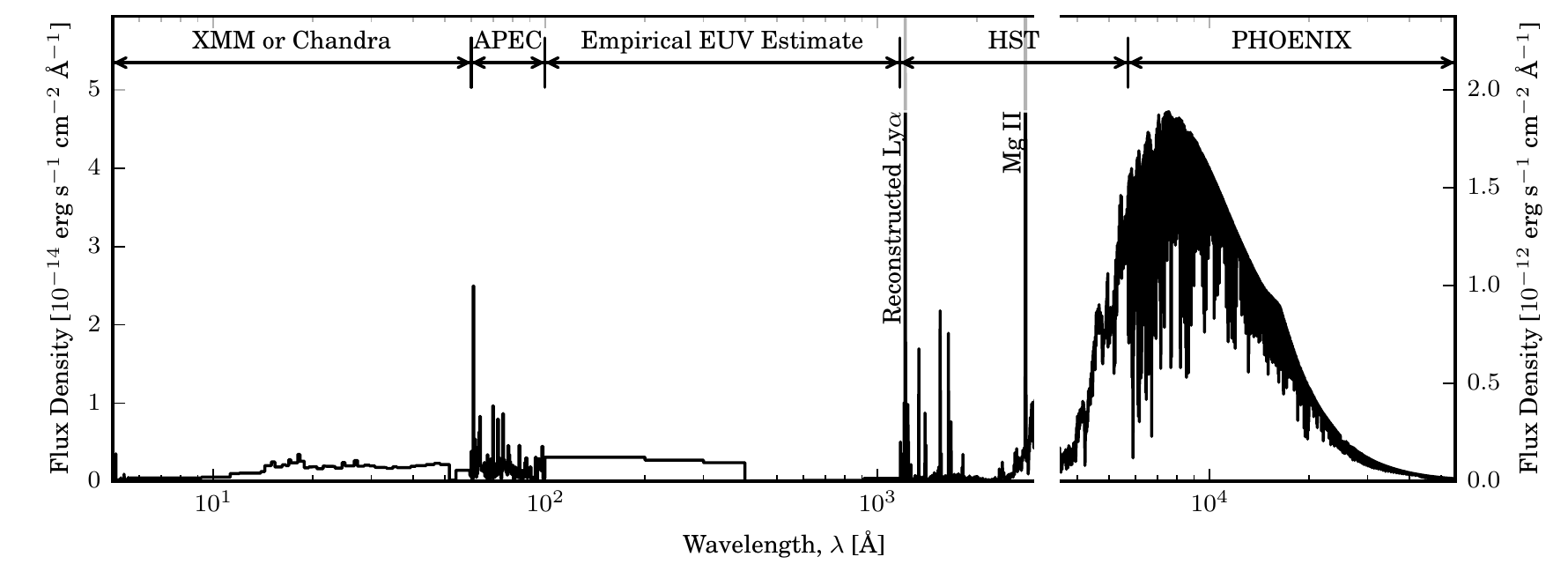}
\caption{Source data ranges for the MUSCLES composite panchromatic SEDs shown with the \srcrngstar\ spectrum for reference.\refchng{8}{The gap in the axes indicates a change in the vertical plot scale.} Both the \lya\ and \Mgii\ lines extend beyond the vertical range of the plot.}
\label{fig:sourceranges}
\end{figure*}

Assembling panchromatic SEDs for the MUSCLES spectral catalog requires data from many sources, \refchng{5}{including both observations and models.
Observational data was obtained using the space telescopes \scraft{HST}, \scraft{Chandra}, and \scraft{XMM-Newton} through dedicated observing programs.
Model output was obtained from APEC plasma models \citep{smith01}, empirical EUV predictions (Paper II), \lya\ reconstructions (Paper II), and PHOENIX atmospheric models \citep{husser13}.}
The approximate wavelength range covered by each source is illustrated in Figure \ref{fig:sourceranges}. 
We describe each data source and the associated reduction process in order of increasing wavelength below, followed by a discussion of how these sources were joined to create panchromatic SEDs.
Paper I includes further details on the rationale and motivation for the various sources.

Two versions of the SED data product, one that retains the native source resolutions and another rebinned to a constant \panres\ resolution, are available for download through the \href{\hlsplink}{MAST High Level Science Product (HLSP) Archive}\footnote{\hlsplink}. 
We made one exception to retaining native source resolutions by upsampling the broad (100 \AA) bands of the EUV estimates to 1 \AA\ binning to ensure accuracy for users that numerically integrate the spectra using the bin midpoints. 
We note, however, that the most precise integration over any portion of the SED is obtained through multiplying bin widths by bin flux densities and summing.

The detailed format of this data product is described in the readme file included in the archive, permitting small improvements to be made over time and reflected in the living readme document. 
The spectral data products retain several types of relevant information, beyond just the flux density in each bin.
This information is propagated through the data reduction pipeline at each step. 
As a result, every spectral bin of the final SEDs provides (at the time of publication) information on the

\begin{itemize}
\item Bin: wavelength of the edges and midpoint [\AA].
\item Flux Density: measurement and error of the absolute value [\fluxcgs\ \perAA] along with the value normalized by the estimated bolometric flux [\perAA] \refchng{6}{(this estimate is discussed in Section \ref{sec:rjtail})}.
\item Exposure: Modified Julian Day of the start of the first contributing exposure, end of the last contributing exposure, and the cumulative exposure time [s].
\item Normalization: any normalization factor applied to the data prior to splicing into the composite SED (applies only to PHOENIX model and \scraft{HST}-STIS data, see Sections \ref{sec:uvreduction} and \ref{sec:visreduction}).
\item Data source: a bit-wise flag identifying the source of the flux data for the bin.
\item Data quality: bit-wise flags of data quality issues (\scraft{HST} data only).
\end{itemize}

\subsection{X-ray data}

We obtained X-ray data with \scraft{XMM-Newton} for GJ 832, HD 85512, HD 40307, and \epseri\ and with \scraft{Chandra} for GJ 1214, GJ 876, GJ 581, GJ 436, and GJ 176. 
Complete details of the X-ray data reduction and an analysis of the results will be presented in a follow-on paper (\citealt{brown16}; in prep.). 
Here we provide a brief outline of the reduction process. 
The technique used to observe X-ray photons imposes some challenges to  extracting the source spectrum from the observational data.
The X-ray emission was observed using CCD detectors that have much lower spectral resolution ($\lambda / \Delta \lambda \sim$ 10-15 for the typical 1 keV photons observed from the MUSCLES stars) than the spectrographs used to measure the UV and visible spectra. 
The data were recorded in instrumental modes designed to allow the efficient rejection of particle and background photon events. 
CCD X-ray detectors have very complex instrumental characteristics that are energy-dependent and also depend critically on where on the detector a source is observed.
Broad, asymmetric photon energy contribution functions result from these effects, and it is nontrivial to associate a measured event energy to the incident photon energy. 
Consequently, simply assigning to each detection the most likely energy of the photon that created the detected event and binning in energy does not accurately reproduce the source spectrum.

To estimate the true source spectrum, we used the XSPEC software package \citep{arnaud96} to forward model and  parameterize the observed event list, using model spectra generated by APEC (Astrophysical Plasma Emission Code; \citealt{smith01}). 
For each spectrum, we experimented with fits to plasma models containing one to three temperature components and with elemental abundance mixes that were either fixed to solar values or free to vary (but normally only varying the Fe abundance, because it is the dominant emission line contributor) until we achieved the most reasonable statistical fit. The complexity of the model fit is strongly controlled by the number of detected X-ray events. The resulting model parameters are listed in Table \ref{tbl:xfitparams}.

We then computed the ratio of the number of photons incident in each energy bin to the number actually recorded, and corrected the measured photon counts accordingly. 
The spectral bin widths are variable because the data were binned to contain an equal number of counts per energy bin.

\begin{deluxetable}{lccc}
\tablewidth{0pt}
\tablecaption{Parameters of the APEC model fits to the X-ray data. \label{tbl:xfitparams}}

\tablehead{\ch{Star} & \ch{$kT$\tnm{a}} & \ch{$EM_{i}/EM_1$\tnm{b}} & \ch{$F_X$\tnm{c}}\\
 & \ch{[keV]} &  & \ch{[$10^{-14}$ erg s$^{-1}$ cm$^{-2}]$}}
		    
\startdata
GJ 1214 & 0.2\tnm{d} &  & $\leq 0.11$ \\
GJ 876 & $0.80 \pm 0.14$ &   & $9.1 \pm 0.8$ \\
 & $0.14 \pm 0.04$ & $3.5 \pm 4.6$ & \\
GJ 436 & $0.39 \pm 0.03$ &  & $1.2 \pm 0.1$ \\
GJ 581 & $0.26 \pm 0.02$ &  & $1.8 \pm 0.2$ \\
GJ 667C & $0.41 \pm 0.03$ &  & $3.9 \pm 0.3$ \\
GJ 176 & $0.31 \pm 0.02$ &  & $4.8 \pm 0.3$ \\
GJ 832 & $0.38_{-0.07}^{+0.11}$ &  & $6.2_{-0.7}^{+0.8}$ \\
 & $0.09_{-0.09}^{+0.02}$ & $4.8_{-2.9}^{+2.6}$ & \\
HD 85512 & $0.25_{-0.03}^{+0.04}$ &  & $1.9_{-0.3}^{+0.4}$ \\
HD 40307 & $0.15 \pm 0.06$ &  & $1.0\pm0.2$ \\
HD 97658\tablenotemark{e} & \nodata & \nodata & \nodata \\
\epseri & $0.70_{-0.07}^{+0.04}$ &  & $940\pm10$ \\
 & $0.32\pm0.01$ & $3.6_{-0.9}^{+0.7}$ & \\
 & $0.12_{-0.02}^{+0.09}$ & $2.4_{-1.0}^{+2.0}$ & \\
\enddata

\tablenotetext{a}{Plasma temperature. Multiple values for an entry represent multiple plasma components in the model.}
\tablenotetext{b}{For multi-component plasmas, this represents the ratio of the emission measure of the $i^\mathrm{th}$ component to the first component according to the order listed in the $kT$ column. }
\tablenotetext{c}{Flux integrated over full instrument bandpass.}
\tablenotetext{d}{Estimate is not well constrained and is based on an earlier \scraft{XMM-Newton} observation. See Section \ref{sec:exceptions}.}
\tablenotetext{e}{No observations. See Section \ref{sec:exceptions}.}

\end{deluxetable}

The X-ray data do not extend to the start of the EUV region, and this 
results in an energy distribution  
gap that, depending on the S/N of the X-ray observations, starts 
at 30~--~60~\AA\ and ends where the empirical EUV flux estimates \mychng{(see Section \ref{sec:lyareduction})} 
begin at 100~\AA. We filled this gap with the output of the same best-fit 
APEC model that yielded the count-rate corrections for the measured 
X-ray spectrum.

\subsection{EUV and \lya }
\label{sec:lyareduction}

Much of the EUV radiation emitted by a star, along with the core of the \lya\ line, cannot be observed because of absorption and scattering by hydrogen in the interstellar medium. 
However, emission in the wings of the \lya\ line does reach Earth because multiple scatterings in the stellar atmosphere produce broad emission wings that are not affected by the narrower absorption profile produced by \Hi\ and \ion{D}{1} in the ISM.
This allows the full line to be reconstructed by modeling these processes.

\citeauthor{youngblood16} (\citeyear{youngblood16}; Paper II) have reconstructed the intrinsic \lya\ profile from observations made with \scraft{HST} STIS G140M and (for bright sources) E140M data. The narrow slits used to collect these data (52x0.1\arcsec\  for G140M and 0.2x0.06\arcsec\ for E140M) produce a spectrum where the diffuse geocoronal \lya\ ``airglow'' emission is spatially extended beyond the target spectrum and spectrally resolved.
This allows the geocoronal spectrum to be measured and subtracted from the observed spectrum, leaving only the target flux.
The modeling procedure used to reconstruct the intrinsic \lya\ line from the airglow-subtracted data is described in detail in Paper II.
This reconstruction covers \vlyaCutLo\ to \lyaCutHi\ in the MUSCLES spectra.

The EUV is estimated from the intrinsic \lya\ flux for each source using the empirical fits of \cite{linsky14}. 
This also is detailed in Paper II. 
We use these estimates to fill all of the EUV as well as \mychng{the portion of the FUV below \cosCutLo\ where the reflectivity of the  Al+MgF$_2$ coatings in the \scraft{HST} spectrograph optics rapidly declines.}

\subsection{FUV through Blue-Visible}
\label{sec:uvreduction}

\begin{figure*}
\centering
\includegraphics{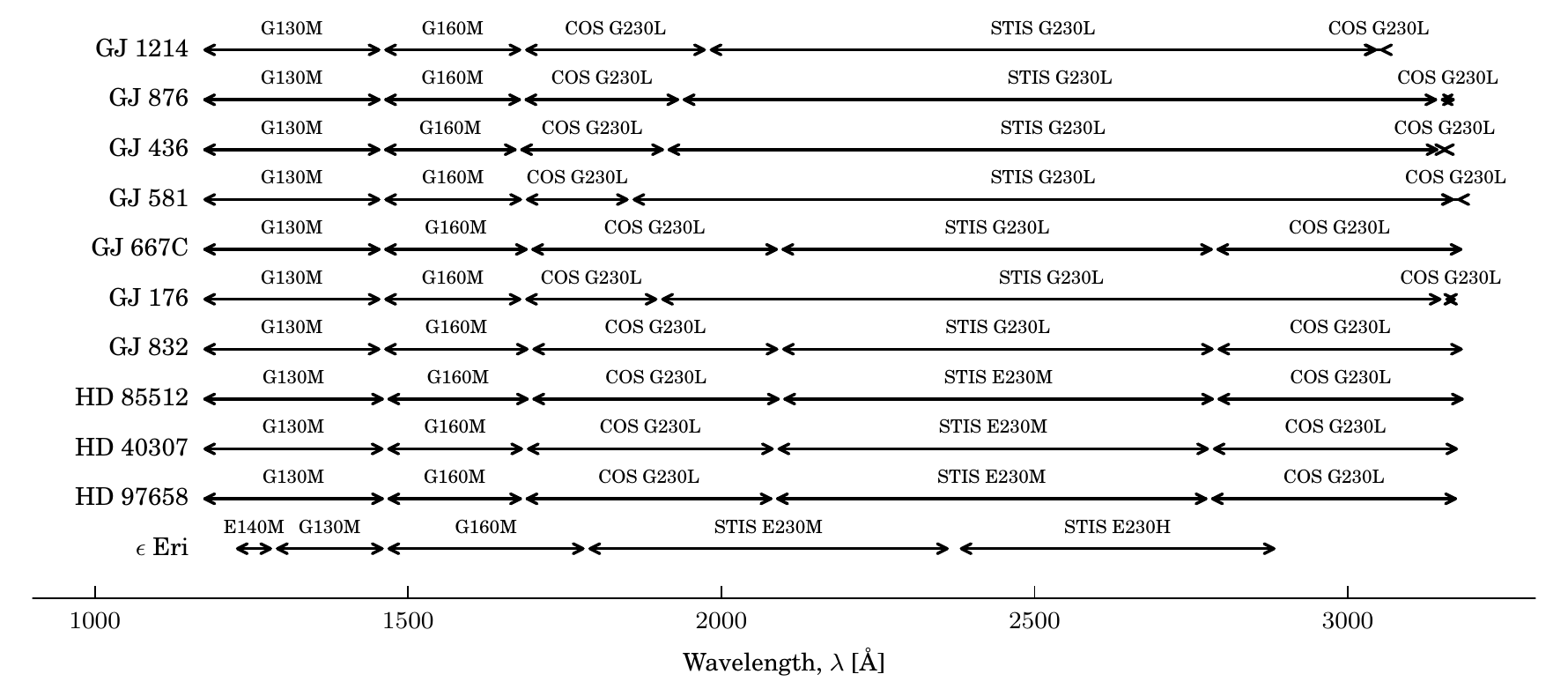}
\caption{Sources of the UV data for the composite SED of each MUSCLES star. Due to space constraints, the labels E140M, G130M, and G160M are used in place of STIS E140M, COS G130M, and COS G160M. \mychng{Note that the STIS G140M or E140M data obtained for each star do not appear on this figure because they were only used to reconstruct the \lya\ emission (Section \ref{sec:lyareduction}) and fill small gaps left by airglow removal (Section \ref{sec:uvreduction}).}}
\label{fig:hstsources}
\end{figure*}

Currently, \scraft{HST} is the only observatory that can obtain spectra at UV wavelengths. 
We used \scraft{HST} with the COS and STIS spectrographs to obtain UV data of all 11 sources.
Obtaining full coverage of the UV required multiple observations using complementary COS and STIS gratings.
The choice of gratings depended on the target brightness.
Figure \ref{fig:hstsources} illustrates these configurations, depicting which instrument and which grating provided the data for the different pieces of each star's UV dataset.
Along with the UV observations, we also obtained a visible spectrum with STIS G430L covering visible wavelengths up to 5700 \AA\ (with the exception of \epseri, for which instrument brightness limits required the use of STIS G430M covering $\sim$3800~--~4075~\AA) since the required observing time once the telescope was already pointed was negligible.

For observations with multiple exposures, we coadded data from each exposure. 
We did the same for the (overlapping) orders of echelle spectra. 
This produced a single spectrum for each instrument configuration.

We inspected all the coadded spectra and culled any data suspected of detector edge effects. 
We also removed emission from the geocoronal airglow present in the COS G130M spectra if the airglow line was visible in the spectrum of at least one \refchng{10}{MUSCLES source. 
The wavelengths at which we inspected the MUSCLES data for this emission are those listed in the \href{http://www.stsci.edu/hst/cos/calibration/airglow\_table.html}{MAST database of airglow lines}\footnote{http://www.stsci.edu/hst/cos/calibration/airglow\_table.html} compiled from airglow-only observations.}
The removed lines were \Ni\ $\lambda$1134, \Hei\ $\lambda$584 at second order (1168 \AA), \Ni\ $\lambda$1200, \Oi\ $\lambda$1305, and \Oi\ $\lambda$1356. 
The resulting gaps were later filled with STIS data where there was overlap and a quadratic fit to the nearby continuum where there was not. 

\subsubsection{A Discrepancy in the Absolute Level of Flux Measurements by COS and STIS}
\label{sec:absflux}

Several instrument configurations of the \scraft{HST} data have overlapping wavelength ranges. 
This allowed us to compare the fluxes from these configurations prior to stitching the spectra into the final panchromatic SED. 
Where there was sufficient signal for a meaningful comparison, STIS always measured lower fluxes than COS by factors of 1.1~--~2.4. 
The cause of this discrepancy could be a systematic inaccuracy in the STIS data, the COS data, or both. 

The STIS G430L data can be compared against external data in search of a systematic trend because the grating bandpass overlaps with the standard B band for which many ground-based photometric measurements are available.
Carrying out this comparison showed that the fluxes measured with the STIS G430L grating were lower than ground based B-band photometry for every star.
The magnitude of these discrepancies varied from the discrepancies between STIS and COS data, but not beyond what is reasonable given uncertainties in the B-band photometry.
A plausible cause of systematically low flux measurements by STIS could be imperfect alignment of the spectrograph slit on the target.
This can produce significant flux losses when a narrow slit is used (\citealt{biretta15}; see Section 13.7.1).
No such comparison with external data was possible for the COS data.
While there is overlap of COS data with \scraft{GALEX} bands, \scraft{GALEX} photometry is only available for roughly half the targets and uncertainties are very large. 

\begin{figure}
\centering
\includegraphics{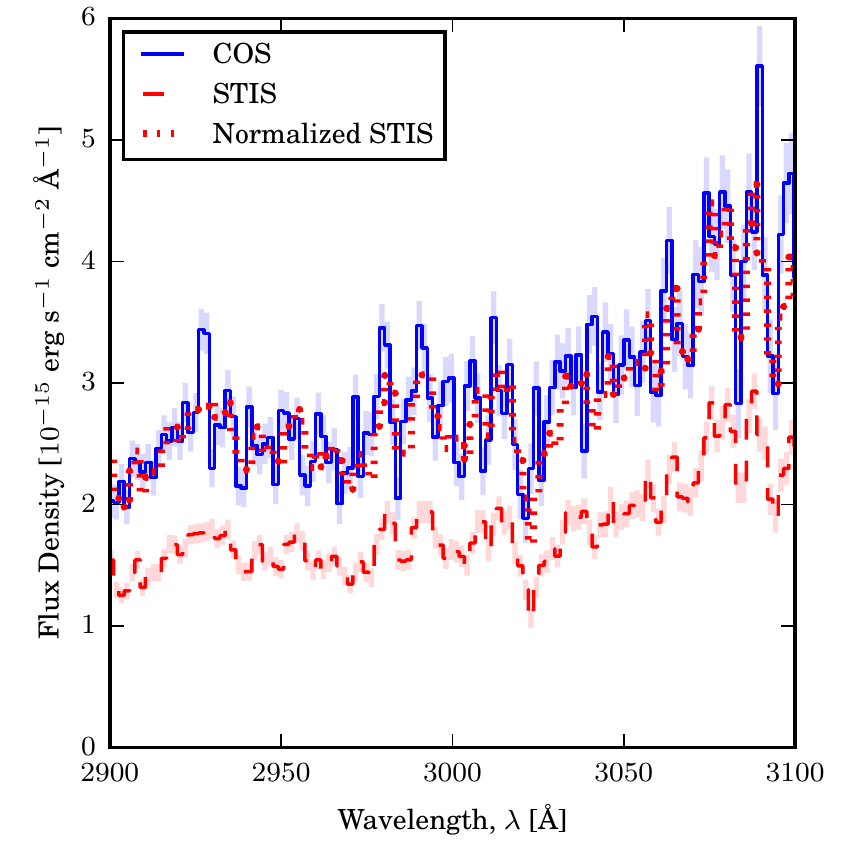}
\caption{An example of normalizing STIS to COS data. The figure shows a portion of the NUV data for GJ 436 from the STIS G230L and COS G230L observations, identically binned for comparison. The COS flux exceeds the STIS flux by a wide margin. Normalizing the STIS data in lieu of possible slit losses produces good agreement.}
\label{fig:normexample}
\end{figure}

Given the low STIS fluxes relative to ground-based photometry, the plausible explanation of such low fluxes, and the lack of an external check for the COS absolute flux accuracy, we chose to treat the absolute flux levels to be accurate for COS and inaccurate for STIS. 
Thus, for each STIS spectrum, we normalized to overlapping COS data whenever the difference in flux in regions of high signal was sufficiently large to admit a $<5\%$ false alarm probability.
This condition was met for all of the G230L and E230M spectra (excepting \epseri\ for which comparable COS NUV data could not be collected due to overlight concerns), a few G140M spectra, and the E140M spectrum of \epseri. 
The normalization factor was simply computed as the ratio of the integrated fluxes in the chosen region.
We normalized data from each instrumental configuration separately, rather than normalizing all STIS data by the same factor, to allow for potential variations in throughput between instrument configurations.
An example of this normalization is displayed in Figure \ref{fig:normexample}.

The STIS G430L spectra have a small amount of overlap with COS G230L near 3100 \AA\ that could be used for normalization. 
However, the G430L data are of poor quality in this region, making normalization factors very uncertain and heavily dependent on the wavelength range used. 
Thus, we normalized the G430L data to all available non-MUSCLES photometry via a PHOENIX model fit, detailed in the following section.

The normalization factors applied are recorded pixel-by-pixel as a separate column in the final data files.

\subsection{Visible through IR}
\label{sec:visreduction}

\begin{deluxetable*}{llrrlrrlrlrl}
\tablewidth{0pt}
\tabletypesize{\scriptsize}
\tablecaption{Selected properties of the stars in the sample. \label{tbl:starprops}} 

\tablehead{ \colhead{Star} & \colhead{Type} & \colhead{$d$} & \colhead{V} & \colhead{ref} & \colhead{MUSCLES $T_{eff}$} & \colhead{Literature $T_{eff}$} & \colhead{ref} & \colhead{$\log{g}$} & \colhead{ref} & \colhead{$\left [ Fe/H \right ] $} & \colhead{ref} \\ \colhead{} & \colhead{} & \colhead{[pc]} & \colhead{} & \colhead{} & \colhead{[K]} & \colhead{[K]} & \colhead{} & \colhead{[[cm s$^{-2}$]} & \colhead{} & \colhead{} & \colhead{} }
\startdata
GJ 1214 & M4.5 & 14.6 & $ 14.68 \pm 0.02 $ & 1 & $ 2935 \pm 100 $ & $ 2817 \pm 110 $ & 2 & $ 5.06 \pm 0.52 $ & 3 & $ 0.05 \pm 0.09 $ & 2\\
GJ 876 & M5 & 4.7 & $ 10.192 \pm 0.002 $ & 4 & $3062_{-130}^{+120}$ & $ 3129 \pm 19 $ & 5 & $ 4.93 \pm 0.22 $ & 3 & $ 0.14 \pm 0.09 $ & 2\\
GJ 436 & M3.5 & 10.1 & $ 10.59 \pm 0.08 $ & 6 & $ 3281 \pm 110 $ & $3416_{-61}^{+54}$ & 7 & $ 4.84 \pm 0.16 $ & 3 & $ -0.03 \pm 0.09 $ & 2\\
GJ 581 & M5 & 6.2 & $ 10.61 \pm 0.08 $ & 6 & $ 3295 \pm 140 $ & $ 3442 \pm 54 $ & 8 & $ 4.96 \pm 0.25 $ & 3 & $ -0.20 \pm 0.09 $ & 2\\
GJ 667C & M1.5 & 6.8 & 10.2 & 9 & $ 3327 \pm 120 $ & $ 3445 \pm 110 $ & 2 & $ 4.96 \pm 0.25 $ & 3 & $ -0.50 \pm 0.09 $ & 2\\
GJ 176 & M2.5 & 9.3 & 10.0 & 10 & $ 3416 \pm 100 $ & $ 3679 \pm 77 $ & 5 & $ 4.79 \pm 0.13 $ & 3 & $ -0.01 \pm 0.09 $ & 2\\
GJ 832 & M1.5 & 5.0 & 8.7 & 10 & $ 3816 \pm 250 $ & $ 3416 \pm 50 $ & 11 & $ 4.83 \pm 0.15 $ & 3 & $ -0.17 \pm 0.09 $ & 2\\
HD 85512 & K6 & 11.2 & 7.7 & 10 & $4305_{-110}^{+120}$ & $ 4400 \pm 45 $ & 12 & $ 4.4 \pm 0.1 $ & 12 & $ -0.26 \pm 0.14 $ & 12\\
HD 40307 & K2.5 & 13.0 & 7.1 & 10 & $ 4783 \pm 110 $ & $ 4783 \pm 77 $ & 12 & $ 4.42 \pm 0.16 $ & 12 & $ -0.36 \pm 0.02 $ & 12\\
HD 97658 & K1 & 21.1 & 7.7 & 10 & $ 5156 \pm 100 $ & $ 5170 \pm 50 $ & 13 & $ 4.65 \pm 0.06 $ & 14 & $ -0.26 \pm 0.03 $ & 14\\
$\mathrm{\epsilon}$ Eri & K2.0 & 3.2 & 3.7 & 15 & $ 5162 \pm 100 $ & $ 5049 \pm 48 $ & 12 & $ 4.45 \pm 0.09 $ & 12 & $ -0.15 \pm 0.03 $ & 12\\
\enddata

\tablerefs{(1) \citealt{weis96}; (2) \citealt{neves14}; (3) \citealt{santos13}; (4) \citealt{landolt09}; (5) \citealt{braun14}; (6) \citealt{hog00}; (7) \citealt{braun12}; (8) \citealt{boyajian12}; (9) \citealt{mermilliod86}; (10) \citealt{koen10}; (11) \citealt{houdebine10}; (12) \citealt{tsantaki13}; (13) \citealt{grootel14}; (14) \citealt{valenti05}; (15) \citealt{ducati02}}

\tablecomments{We located many of these parameters through the PASTEL \citep{soubiran10} catalog, but have provided primary references in this table.}

\end{deluxetable*}

We used synthetic spectra generated by \cite{husser13} from a PHOENIX stellar atmosphere model to fill the range from the blue visible at $\sim$5700 \AA\ out to 5.5 \micron\ where their model spectra truncate. 
The choice to use model output instead of observations enabled greater consistency in the treatment of the visible and IR between sources, given that for some sources we were unable to acquire optical and IR spectra within the same time window as the X-ray and UV observations.
The  visible and IR emission of low-mass stars is well reproduced by PHOENIX models.
The \cite{husser13} PHOENIX spectra cover a grid in effective temperature (\teff), surface gravity (\logg), metallicity (\feh), and $\alpha$ metallicity. 
The $\alpha$ metallicity is used to specify the abundance of the elements O, Ne, Mg, Si, S, Ar, Ca, and Ti relative to Fe.
However, we found no data on $\alpha$ metallicity for the target stars, so we treated each as having the solar value. For \teff, \logg, and \feh, we found literature values for all stars, as listed in Table \ref{tbl:starprops}. 

\refchng{13}{
The PHOENIX model output provided by \cite{husser13} has an arbitrary scale.
Thus, the output must be normalized to match the absolute flux level of the star.
To constrain this normalization using data on the star, we collected all external photometry for the star returned by the VizieR Photometry Viewer\footnote{http://vizier.u-strasbg.fr/vizier/sed/} within a 10" search radius.
An exception was GJ 667C, where companion stars A and B prevented a position-based search. 
Instead, we collected photometry specifically associated with the object from the Denis catalog, the UCAC4 catalog, and the HARPS survey.
For all other stars, we verified that no other sources fell within the search radius in 2MASS imagery.
References for the photometry we collected are given in Table \ref{tbl:photrefs}.}

\refchng{13}{
For several stars, simply retrieving a PHOENIX spectrum from the \citet{husser13} grid using stellar parameters found in the literature produced a spectrum with an overall shape that did not match the collected photometry.
The overall spectral shape is primarily driven by \teff, so, to correct this mismatch, we wrote an algorithm to search the \cite{husser13} grid of PHOENIX spectra for the best-fit \teff.
Our fitting algorithm operated by taking a supplied \teff\ and the literature values for \logg\ and \feh\ and tri-linearly interpolating a spectrum from the \cite{husser13} grid.
It then computed the best-fit normalization factor that matched the spectrum to stellar photometry.}
The normalization factor was computed analytically via a min-\chisquare\ fit to the photometry under the assumption of identical S/N for each point.
We estimated the S/N as the RMS of the normalized residuals, that is

\begin{equation}
\frac{F_{o,i}}{\sigma_i} = \left[ \frac{1}{N}\sum_i{ \frac{\left(F_{o,i} - F_{c,i}\right)^2}{F_{o,i}^2}}\right]^{-1/2},
\end{equation}

where $\sigma_i$ is the uncertainty on the observed flux $F_{o,i}$, $F_{c,i}$ is the synthetic flux computed by applying the transmission curve of the filter used to measure $F_{o,i}$ to the PHOENIX model, and $i$ indexes the $N$ available flux measurements. 
Estimating measurement uncertainties permits the inclusion of data lacking quoted uncertainties and mitigates possible underestimation of uncertainties in those data for which they are given.
\refchng{12}{Uncertainties on the best-fit normalization factors span \vphxnormminerr\% (\phxnormminerrstar) to \vphxnormmaxerr\% (\phxnormmaxerrstar) with a median uncertainty of \vphxnormmederr\%, and the uncertainty in normalization correlates well with the target V magnitude.}

After normalization, the algorithm checked for outlying photometry by computing the deviation for which the false alarm probability of a point occurring beyond the deviation was $<$ \normPhotOutlierCut.
Points beyond that deviation were culled and the fit recomputed to convergence.
Once converged, the algorithm computed the likelihood of the model given the data.
This then permitted a numerical search for the maximum-likelihood \teff.

Once the best-fit \teff\ was found, the algorithm sampled the likelihood function to find the 68.3\% confidence interval on \teff.
For this search, the photometry was fixed to the outlier-culled list and associated S/N estimate from the best fit.
\refchng{13}{This confidence interval provides a statistical uncertainty estimate, but the assumptions made in the fitting process (PHOENIX model, constant S/N estimated from residuals) introduces a further, systematic uncertainty. 
We estimated this systematic uncertainty to be \teffSystemErr\ and added it in quadrature to the statistical uncertainty to produce a final uncertainty estimate for the \teff\ values we computed for each star.}

\begin{deluxetable}{ll}
\tablewidth{\columnwidth}
\tabletypesize{\scriptsize}
\tablecaption{References for stellar photometric measurements. \label{tbl:photrefs}}

\tablehead{\colhead{Star} & \colhead{Photometry References}}
\startdata
GJ 1214 & 1,2,3,4,5,6,7,8,9\\
GJ 876 & 10,11,12,7,13,6,14,15,16,3,17,18,19,20,9\\
GJ 436 & 21,22,12,2,7,13,14,3,17,6,19,23,9\\
GJ 581 & 10,21,1,24,7,13,17,12,14,19,23,9\\
GJ 667C & 25,26,27\\
GJ 176 & 11,22,7,13,6,14,15,16,12,18,19,9\\
GJ 832 & 28,10,29,30,7,13,14,16,31,32,23,19,33,8,9\\
HD 85512 & 28,34,35,36,37,30,23,13,14,31,18,32,33,8,9\\
HD 40307 & 28,34,35,11,37,30,23,13,9,3,18,32,33,8,38\\
HD 97658 & 11,37,30,3,13,39,31,18,32,33,8,9\\
$\mathrm{\epsilon}$ Eri & 35,37,30,40,41,13,14,42,9,18,32,33,8,43\\
\enddata

\tablerefs{(1) \citealt{2004AAS...205.4815Z}; (2) \citealt{2014MNRAS.444..711T}; (3) \citealt{2013A&A...556A.150S}; (4) \citealt{2011PASP..123..412W}; (5) \citealt{2014yCat.2328....0C}; (6) \citealt{2012ApJ...748...93R}; (7) \citealt{2011AJ....142..138L}; (8) \citealt{2012yCat.2311....0C}; (9) \citealt{2003yCat.2246....0C}; (10) \citealt{2015AJ....149....5W}; (11) \citealt{2006ApJ...638.1004A}; (12) \citealt{2012A&A...546A..61D}; (13) \citealt{2008A&A...488..401R}; (14) \citealt{2003ApJ...582.1011S}; (15) \citealt{2004AJ....127.3043Z}; (16) \citealt{2013A&A...549A.109B}; (17) \citealt{2009ApJ...704..975J}; (18) \citealt{2010PASP..122.1437P}; (19) \citealt{2014MNRAS.443.2561G}; (20) \citealt{2014AJ....148..119F}; (21) \citealt{2010AJ....139.2440R}; (22) \citealt{2004AJ....128..463R}; (23) \citealt{2015A&C....10...99A}; (24) \citealt{2009ApJ...705.1226B}; (25) \citealt{2005yCat.2263....0T}; (26) \citealt{2013A&A...553A...8D}; (27) \citealt{2012yCat.1322....0Z}; (28) \citealt{2011AJ....142...15G}; (29) \citealt{2013A&A...551A..36N}; (30) \citealt{2007A&A...474..653V}; (31) \citealt{2015yCat.5145....0M}; (32) \citealt{2012AstL...38..331A}; (33) \citealt{2001KFNT...17..409K}; (34) \citealt{2009ApJ...705...89L}; (35) \citealt{2009A&A...501..941H}; (36) \citealt{2012ApJS..200...15A}; (37) \citealt{2010A&A...515A.111S}; (38) \citealt{2013A&A...555A..11E}; (39) \citealt{2011MNRAS.411..435B}; (40) \citealt{2002A&A...384..180F}; (41) \citealt{2009ApJS..184..138H}; (42) \citealt{2010AJ....139.1242A}; (43) \citealt{1879RNAO....1.....G}}

\end{deluxetable}

\begin{figure*}
\centering
\includegraphics{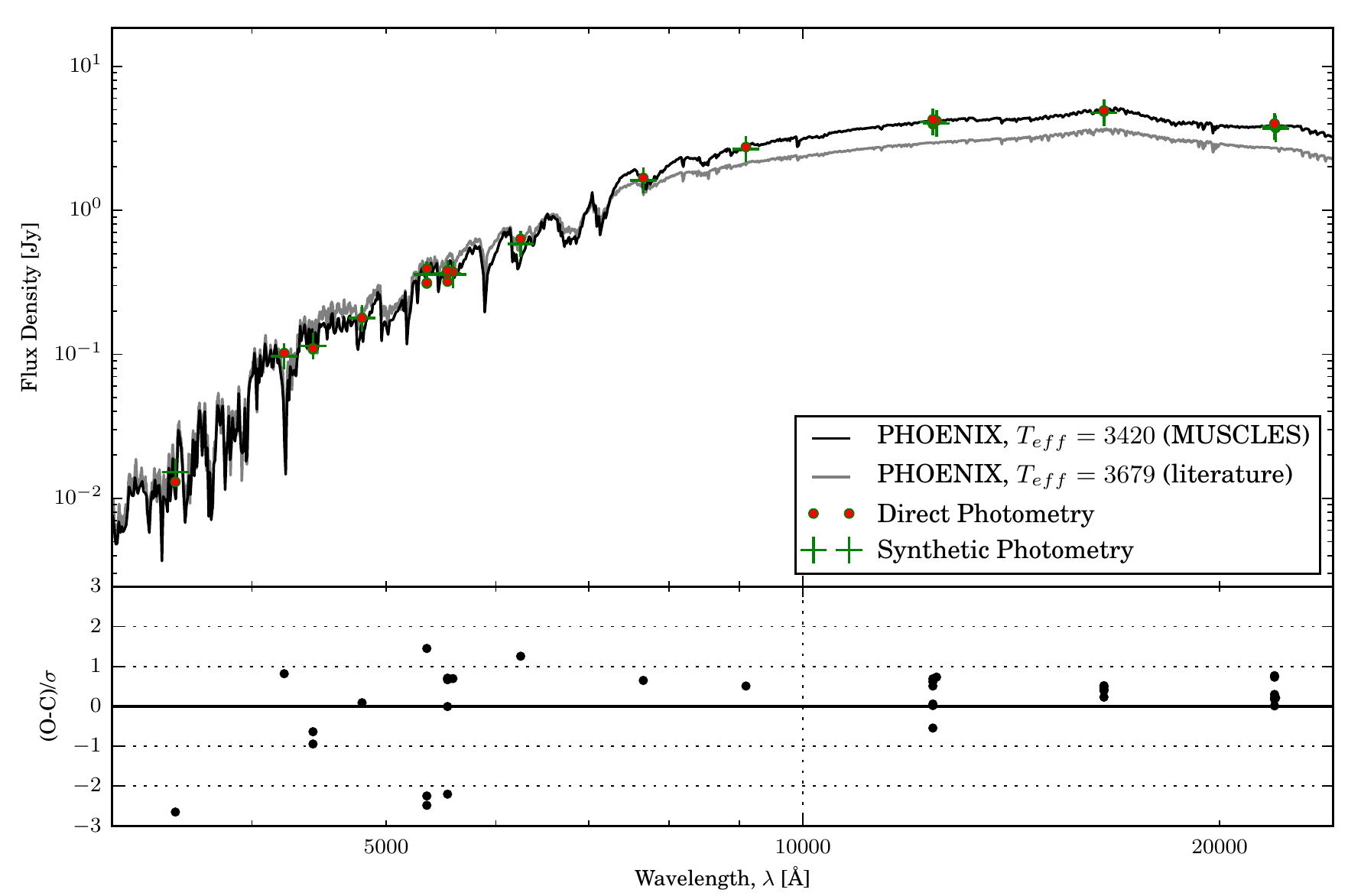}
\caption{The normalization and \teff\ fit of PHOENIX model output to non-MUSCLES photometry for \teffFitStar. \refchng{14}{Synthetic photometry (green crosses) computed from the PHOENIX spectrum with best-fit \teff\ (black line) agrees well with all photometric data (red dots). In contrast, the PHOENIX spectrum where \teff\ is fixed to the literature value (gray line) produces a noticeably poor fit to the photometry. Photometric data are plotted at the mean filter wavelength.}}
\label{fig:Teff_fit}
\end{figure*}

Our best-fit \teff\ values and corresponding literature values are listed in Table \ref{tbl:starprops}. 
Figure \ref{fig:Teff_fit} illustrates the discrepancy between the literature \teff\ values and the photometry for the worst case. 
The figure includes one fit computed with \teff\ as a free parameter and another with \teff\ fixed to the literature value.
The shape of the PHOENIX spectrum interpolated at the literature value does not conform to the photometry.

After finding the best-fit PHOENIX spectrum, we used it to normalize the STIS G430L data according to the ratio of integrated fluxes in the overlap redward of \visPhxNormrangeLo.

\subsection{Creating Composite Spectra}
We spliced together the spectra from each source to create the final spectrum. 
When splicing, observations were always (with a few exceptions mentioned in the next section) given preference over the APEC or PHOENIX models. 
When two observations were spliced, we chose the splice wavelength to minimize the error on the integrated flux in the overlapping region. 
This causes the splice locations to vary somewhat from star to star (see Figure \ref{fig:hstsources}).
Once all of the spectra were joined, we filled the remaining small gaps that resulted from removing telluric lines (along with a slight separation between the E230M and E230H spectra in \epseri) with a quadratic fit to the continuum in a region \gapFitSpan\ times the width of the gap. 
Throughout the process, we propagated exposure times, start date of first exposure, end date of last exposure, normalization factors, data sources, and data quality flags on a pixel-by-pixel basis into the final data product.

\subsection{Special Cases}
\label{sec:exceptions}

The diversity of targets and data sources included in this project necessitated individual attention in the data reduction process. 
In all cases, spectra were separately examined and suspect data, particularly data near the detector edges, were culled. 
In a variety of cases, detailed below, we tweaked the data reduction process.

The \scraft{HST} STIS G430L data for all stars show large scatter with many negative flux bins at the short-wavelength end of the spectrum.
For most stars, this region is small enough to be significantly overlapped by COS G230L data that we have used in the composite SED. 
However, for \textbf{GJ 667C} and \textbf{GJ 1214} this region is much larger than the COS G230L overlap.
In these cases, we culled the STIS G430L data from the long-wavelength edge of the COS G230L spectra to $\sim$\gvisCustomCut, roughly where the negative-flux pixels of the STIS G430L data cease, and filled the gap with the PHOENIX model. 
\mychng{While the PHOENIX models used here omit emission from the stellar upper atmosphere that dominates short-wavelength flux (see Section \ref{sec:badphoenix}), this omission does not begin to have a substantial effect until shortward of where the COS G230L coverage begins in other targets.}

\textbf{GJ 667C} showed fluxes for all STIS data several times below that of the COS data. 
Inspection of the acquisition images revealed the spectrograph slit was very poorly centered on the source.
We did not alter the reduction process for this star, but we note that the STIS data \refchng{15}{were normalized upward by large factors in this case.
Specifically, we computed normalization factors of \Cnormlya\ for the STIS G140M data, \Cnormmgii\ for the STIS G230L data, and \Cnormcont\ for the STIS G430L data.}

The STIS G140M spectra of \textbf{GJ 1214, GJ 832, GJ 581} and  \textbf{GJ 436} and the STIS G230L spectrum of GJ 1214 were improperly extracted by the CalSTIS pipeline because the pipeline algorithm \mychng{could not} locate the spectral trace on the detector. 
However, the spectral data were present in the fluxed 2D image of the spectrum (the .x2d files). 
We extracted spectra for these targets by summing along the spatial axis in a region of pixels centered on the signal in the spectral images with a proper background subtraction and correction for excluded portions of the PSF. 
We checked our method using stars where CalSTIS succeeded in locating and extracting the spectrum and found good agreement.

The \scraft{Chandra} data for \textbf{GJ 1214} provided only an upper limit on the X-ray flux. As a result, we decided to use a previous model fit to \scraft{XMM-Newton} data \citep{lalitha14} to fill the X-ray portion of GJ 1214's spectrum.

We did not acquire X-ray data for \textbf{HD 97658}. 
To fill this region of the spectrum, we use data from HD 85512 \refchng{16}{scaled by the ratio of the bolometric fluxes of the two stars. 
These stars have nearly identical \ion{Fe}{12} $\lambda$1242 emission relative to their bolometric flux. 
The \ion{Fe}{12} ion has a peak formation temperature above 10$^6$ K \citep{dere09}, associating it with coronal emission.
Thus, the similarity in \ion{Fe}{12} emission between HD 97658 and HD 85512 suggests similar levels of coronal activity.
This conclusion is further supported by the similar ages estimated by \cite{bonfanti16} of $9.70\pm2.8$ Gyr for HD 97658 and $8.2\pm3.0$ Gyr for HD 85512.}

Both \textbf{GJ 581} and \textbf{GJ 876} were observed by \scraft{Chandra} at two markedly different levels of X-ray activity. We include in the panchromatic SEDs the observations taken when the stars were less active.

Finally, the spectral image of the STIS G230L exposure of \textbf{GJ 436} shows a faint secondary spectrum separated by about 80 mas from the primary. 
It is very similar in spectral character, if not identical, to the GJ 436 spectrum.
An identical exposure from 2012 does not show the same feature, but this is unsurprising given the high proper motion of GJ 436 \citep{leeuwen07}.
If a second source is a significant contributor, this could impart an upward bias to the GJ 436 flux.

\subsection{Notes on Data Quality}
Users should keep several important characteristics of the panchromatic SEDs in mind when using them.

\subsubsection{Flares}
These stars exhibited flares in the UV, some very large (see Paper I), during the observations. 
The rates of such flares on these stars is mostly unconstrained: there is little or no previous data that could provide a good estimate of whether the observed flares were typical or atypical for the target.
Thus, the safest assumption is to treat the MUSCLES observations as typical.
As such, we did not attempt to separate the data into times of flare and quiescence.
These observations should thus be treated as roughly typical of any average of one to a few hours of UV data for a given star. 
These flares will be the subject of a future publication (\citealt{loyd16b}; in prep).

\subsubsection{Wavelength Calibration}
We observed some mismatches in the wavelengths of spectral features in the NUV data from COS and STIS.
This mismatch is most pronounced at the \Siii\ $\lambda\lambda 1808,1816$ lines, where the line centers in the STIS E230M data for \epseri\ (the only STIS observation that resolves the lines) match the true line wavelengths, but the COS G230L data of all other targets show the lines shifted $\sim$4~\AA\ ($\sim$660~km~\pers) blueward.
This shift in the COS G230L wavelength solution is not present at the next recognizable spectral feature redward of \Siii, \Mgii.
Flux from \Siii\ and \Mgii\ is captured on different spectral ``stripes'' on the COS G230L detector, suggesting that the entire stripe capturing \Siii\ flux might be poorly calibrated in comparison to the stripe capturing \Mgii\ flux.
The \Mgii\ $\lambda\lambda~2796,2803$ lines are in an area of overlap between the COS G230L data and STIS E230M (K stars) and G230L (M stars) data, enabling a direct comparison of the wavelength solutions for these spectra at \Mgii. 
Making this comparison reveals that STIS E230H and COS G230L data agree to within 1 \AA, while STIS G230L and COS G230L data can disagree by up to $~2$~\AA\ ($\sim$330~km~\pers).
In the latter cases, the COS data more closely match the true wavelength of the \Mgii\ lines.

The NUV discrepancy for the narrow-slit STIS G230L observations at \Mgii\ could result from the same target centering issues that may cause the flux discrepancies discussed in Section \ref{sec:absflux}.
We do not understand the source of the erroneous COS wavelength solution at the \ion{Si}{2} lines.

In the FUV, wavelength calibrations between COS G130M and STIS G140M data at the \Nv\ $\lambda\lambda$1238,1242 lines typically agree to within a single pixel width of the STIS G140M data ($\sim$0.05~\AA, 10 km~\pers).

We did not alter the wavelength calibration of any spectra. 
Neither do we attempt to shift the spectra to the rest frame of the star. 
Such a shift would be $<$0.05\% for any target ($<0.5$ \AA\ at 1000 \AA). 
To place the wavelength miscalibrations and target redshifts in context, we note that the molecular cross section spectra used in Section \ref{sec:dissrates} have a resolution of 1 \AA.

\subsubsection{Negative Flux}
Users will notice bins with negative flux density in many low-flux regions in the UV. 
This is a result of subtracting a smoothed background count rate from regions where noise dominates the signal (Section 3.4\footnote{\url{http://www.stsci.edu/hst/stis/documents/handbooks/currentDHB/ch3_stis_calib5.html}} of the STIS Data Handbook, \citealt{bostroem11} and Section 3.4\footnote{\url{http://www.stsci.edu/hst/cos/documents/handbooks/datahandbook/ch3_cos_calib5.html}} of the COS Data Handbook, \citealt{fox15}).
While a negative flux is unphysical, the background subtraction serves to produce an unbiased estimate of the flux when low-flux regions are integrated.
Thus, we left the negative bins in the low-flux regions unaltered.

\subsubsection{The Rayleigh-Jeans Tail}
\label{sec:rjtail}
The truncation of the PHOENIX spectra at 5.5 \micron\ results in the omission of flux that contributes a few percent to the bolometric flux of the stars. 
The bolometric flux estimate determines the relative level at which each wavelength regime contributes to the stellar SED, so accuracy is important.
We therefore compute and include in the data product a bolometric flux value for each target that incorporates the integral of a blackbody fit to the PHOENIX spectrum from the red end of the PHOENIX range at 5.5 \micron\ to $\infty$.
Whenever we present bolometrically normalized fluxes, we use this more accurate value as opposed to the integral of the MUSCLES SED alone.

\subsubsection{\epseri\ STIS E230M/E230H Data Compared with PHOENIX Output}
Unlike the other targets, \epseri\ was too bright to permit the collection of \scraft{HST} COS observations in the NUV, so only \scraft{HST} STIS data, specifically using the E230M and E230H gratings, were collected.
Because of the lack of COS data that was used to correct systematically lower flux levels in the STIS NUV data of other stars (Section \ref{sec:absflux}), we examined the \epseri\ STIS NUV data closely.
This star has a high enough \teff\ for photospheric flux to contribute significantly in the NUV.
Thus, the PHOENIX spectrum for \epseri\ can be meaningfully compared to the observation data in this regime.

The lower envelope of the PHOENIX spectrum matches very well with that of the E230M and E230H data. 
However, some portions of the PHOENIX spectrum show emission features well above that seen in the E230M and E230H spectra.
This results in the integrated flux of the PHOENIX spectrum exceeding that of the observations by \epseriEMmismatch\% for E230M and \epseriEHmismatch\% for E230H.
Because of the agreement in the lower envelopes of the spectra, we conclude that this mismatch is likely caused by inaccuracy in the PHOENIX spectrum rather than in the observations.
At longer wavelengths, the \scraft{HST} STIS G430M observation of \epseri\ covering $\sim$3800~--~4100 \AA\ agrees with the PHOENIX spectrum to \epseriGMmismatch\%, suggesting the accuracy of both the observations and PHOENIX output is good in that regime.

\subsubsection{Transits}
We did not attempt to avoid planet transits during observations of the MUSCLES targets to facilitate the \mychng{observatory} scheduling process.
As a result, some observations overlap with planet transits.
We checked for overlap by acquiring transit ephemerides for all hosts where these ephemerides were well established from the NASA Transit and Ephemeris Service\footnote{\url{http://exoplanetarchive.ipac.caltech.edu/cgi-bin/TransitView/nph-visibletbls?dataset=transits}} and comparing these in-transit time ranges to the time ranges of \scraft{HST} observations. 

GJ 1214b transited during one of the \scraft{HST} observations:
one of the three COS G160M exposures is almost fully within GJ 1214 b's geometric transit.
However, the $\sim1$\%  transit depth of GJ 1214b \citep{carter11}, is insignificant in comparison to the 34\% uncertainty on the integrated G160M flux. 

\mychng{GJ 436b was undergoing geometric transit ingress at the end of the third COS G130M exposure and transit egress at the start of the fourth, according to the \cite{knutson11} ephemeris. 
The G130M exposures were sequential, broken up only by Earth occultations.
Consequently, the last four of the five G130M exposures fall within the range of the extended \lya\ transit that begins two hours prior and lasts at least three hours following the geometric transit and absorbs 56\% of the stellar \lya\ emission \citep{kulow14, ehrenreich15}.}
Thus, these observations should be treated as lower limits to the out-of-transit emission of GJ 436 from ions that might be present in the planet's extended escaping cloud.
The geometric transit depth of GJ 436 is under 1\% \citep{torres08}, so only an extended cloud of ions could effect the G130M spectra by an amount that is significant relative to uncertainties. 
We inspected the G130M data for evidence of transit absorption by measuring the flux of the strongest emission lines as a function of time and found no such evidence.
However, because all exposures may be affected by an extended cloud, the lack of a clear transit dip \mychng{might not be} conclusive.
We intend to explore the COS G130M transit data in greater depth in a future work (\citealt{loyd16a}; in prep). 
The reconstructed \lya\ flux stitched into the panchromatic SEDs is not affected. 
It was created from separate STIS observations that \mychng{occurred} outside of transit.

\section{Discussion}
\label{sec:discuss}

\begin{figure*}
\centering
\includegraphics{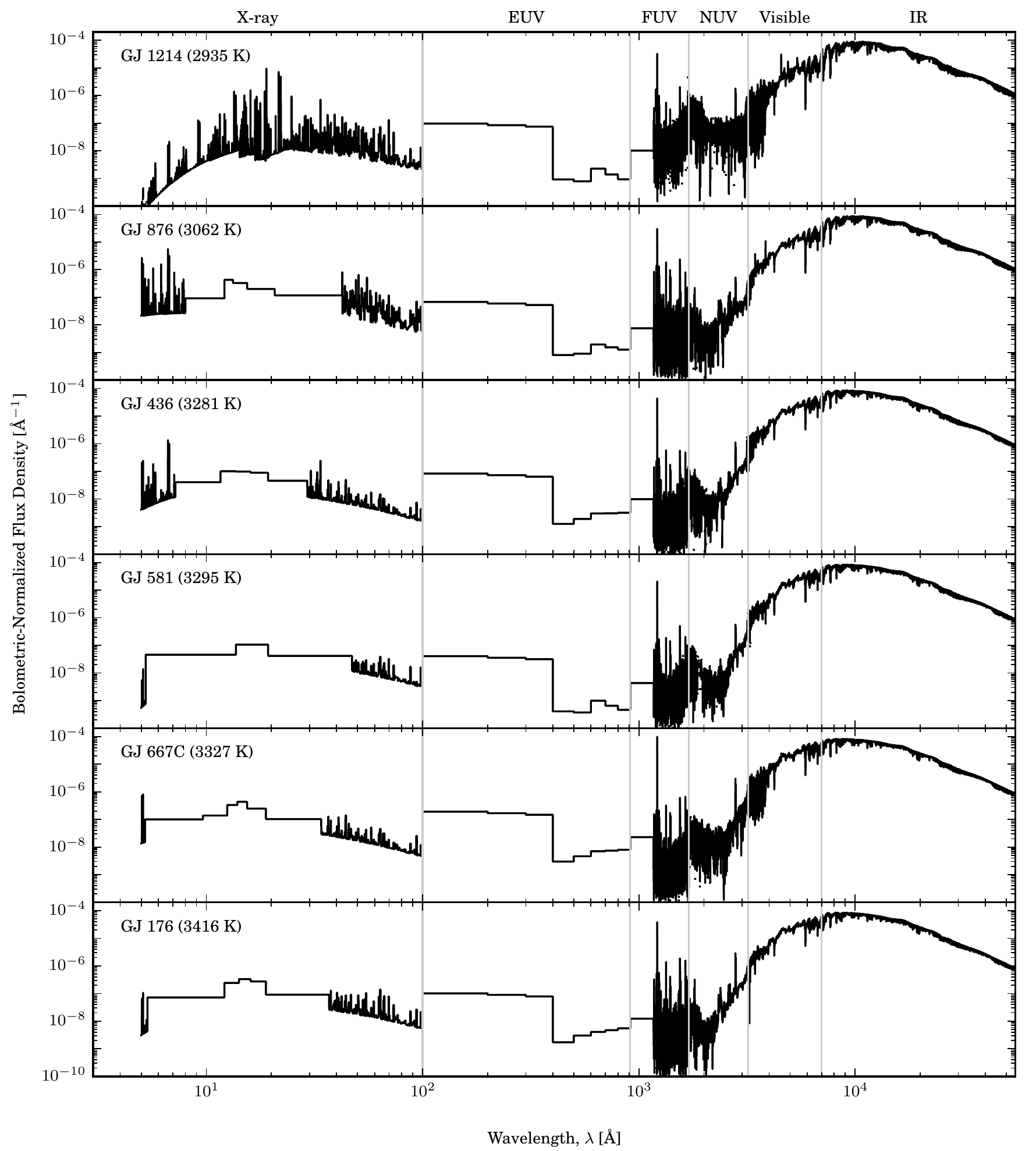} 
\caption{Spectral energy distributions of the MUSCLES stars, normalized by their bolometric flux such that they integrate to unity. 
Axes have identical ranges to facilitate comparisons, each spanning \vstackWmin~--~\vstackWmax~\ustackWmin\ in wavelength and \vstackClip~--~\vstackFmax~\ustackClip\ in flux density. 
\refchng{19}{
Light gray vertical lines show the division between the spectral regions labeled at the top of the plot.
The best-fit effective temperature of each star computed via our PHOENIX model grid search (Section \ref{sec:visreduction})} is listed in parenthesis next to its name.
The normalization enables easy scaling to any star-planet distance; e.g. for an Earth-like planet, multiply by Earth's insolation, \insolationsi.}
\label{fig:spectrastack1}
\end{figure*}

\begin{figure*}
\centering
\includegraphics{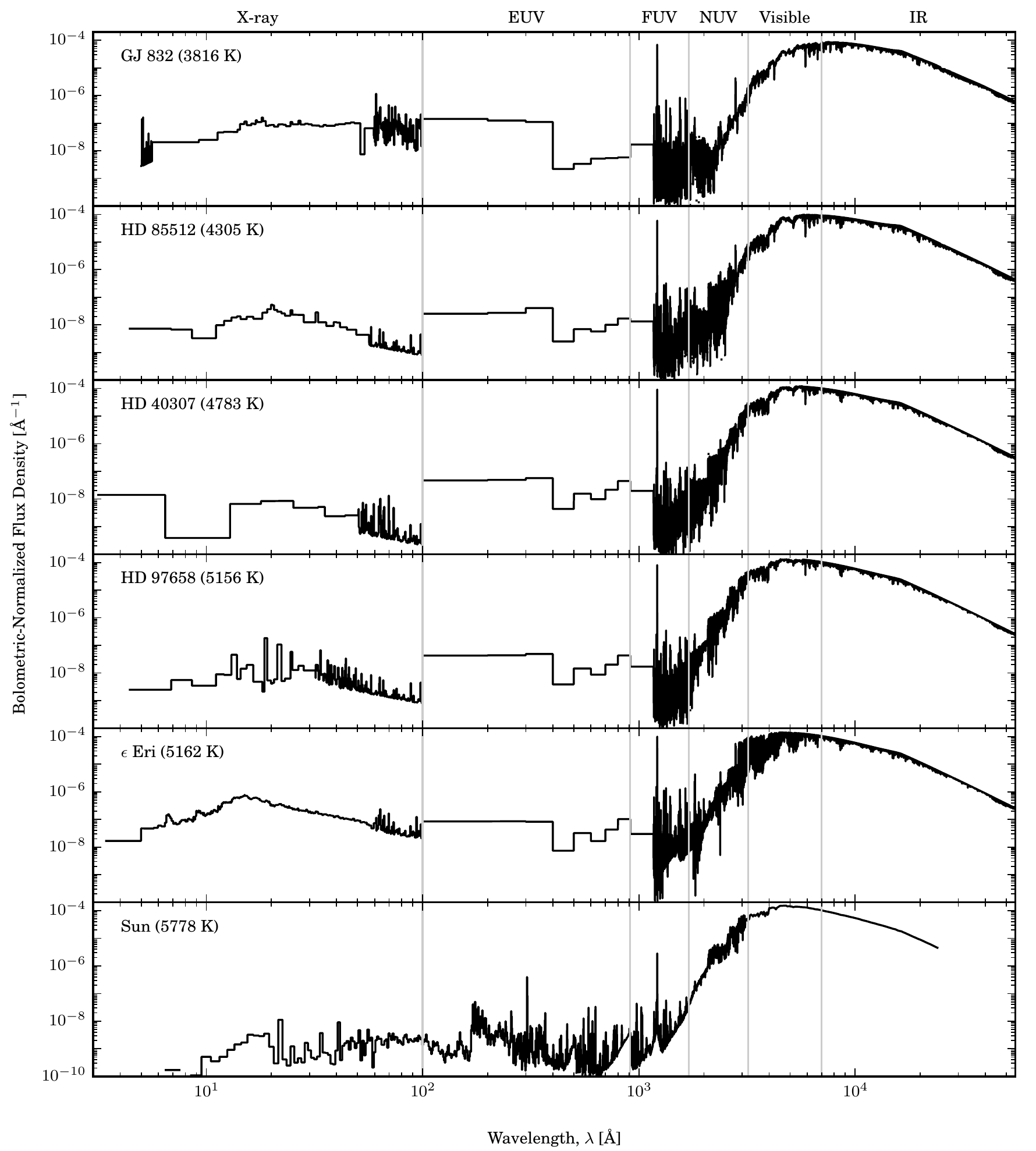}
\caption{Spectral energy distributions of the MUSCLES stars, continued from Figure \ref{fig:spectrastack1}.}
\label{fig:spectrastack2}
\end{figure*}

We display the primary data product of the MUSCLES survey, the panchromatic SEDs for each target, in Figures \ref{fig:spectrastack1} and \ref{fig:spectrastack2}. 
We also include a spectrum of the Sun for comparison to the MUSCLES stars. 
The solar spectrum we present in these figures and in the discussion following is that assembled by \cite{woods09} from several contemporaneous datasets covering a few days of ``moderately active'' solar emission. \refchng{18}{We adopt a value of \insolation\  for the bolometric solar flux at Earth, i.e. Earth's insolation, per resolution B3 of IAU General Assembly XXIX (2015).}

For a detailed discussion of emission line strengths and the correlation of fluxes from lines with different formation temperatures, see Paper I. 
Here we will examine the UV continuum, the degree of deviation from purely photospheric models, and photodissociation rates on orbiting planets.

\subsection{The FUV Continuum}
Hitherto, the sensitivities of FUV instruments, such as the \scraft{International Ultraviolet Explorer} and \scraft{HST} STIS, have rendered observations of anything but isolated emission lines in spectra from low-mass stars challenging.
The greater effective area and lower background rate of \scraft{HST} COS facilitates observations of less prominent spectral features.
\cite{linsky12} used this advantage to study the FUV continua of solar-mass stars, finding that more rapidly rotating stars showed higher levels of FUV surface flux and corresponding brightness temperature. 

In a similar vein, we searched for evidence of an FUV continuum in the spectra of the MUSCLES stars.
Continuum regions were identified by eye through careful examination of the spectra of all 11 targets, similar to the methodology of \cite{france14}, and typically range from \vcontDwMin\ to \contDwMax\ in width. 
The selected regions constitute \contTotalDw\ of continuum spanning \vcontWmin\ ~--~ \contWmax. 
These regions have a bandwidth-weighted mean wavelength of \contMeanW. 
We integrated the flux in each of these bands to create continuum spectra for the targets.

\begin{figure*}
\centering
\includegraphics{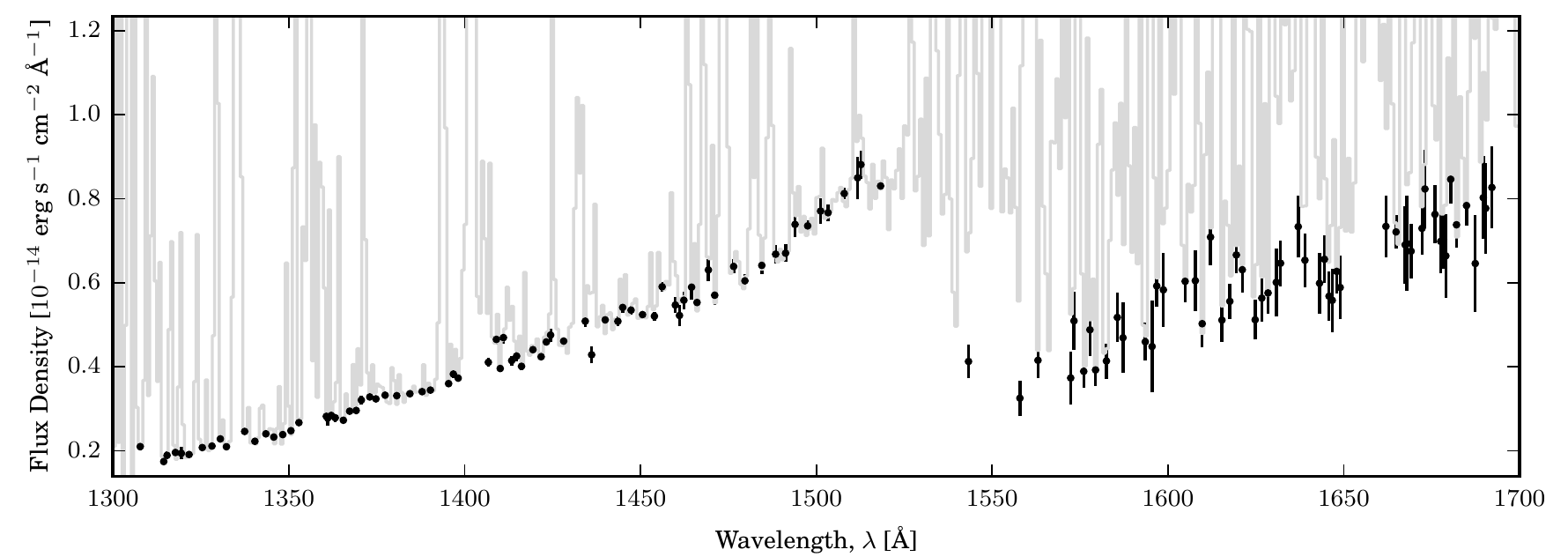}
\caption{Continuum of \epseri, the brightest source with the clearest continuum. Each black point is the average flux density in a continuum band. The width of each band is roughly the size of the point or smaller. The full spectrum (i.e. non-continuum) is plotted in gray in the background, rebinned to $R = \fuvContPlotPow$ for display. The edge occurring between 1500~--~1550 \AA\ is consistent with recombination of \Siii\ to \Sii\ at $\sim$1521 \AA.} 
\label{fig:fuvcont}
\end{figure*}

The continuum spectrum of the brightest target, \epseri, is displayed in Figure \ref{fig:fuvcont}. 
This continuum shows a shape that strongly suggests a recombination edge occurring between 1500 and 1550 \AA. 
This is consistent with the $\sim$1521 \AA\ limit of \Siii\ recombination to \Sii. 
Models of the solar chromosphere have predicted this edge, but it is not observed in the solar FUV spectrum \citep{fontenla09}, nor was it observed in the continua of solar-mass stars studied by \cite{linsky12}.

Data from other targets did not have sufficient S/N to show a clear continuum shape. 
However, by integrating the flux in all \contTotalDw\ of continuum bands, some level of continuum was detected at $>3\sigma$ significance for \vnContDetections/11 targets.
Note that the negative values in Table \ref{tbl:fuvcont} are not concerning given their large error bars. 
A simple detection of continuum emission does not indicate whether this emission is significant relative to line emission in the FUV.
To quantify this significance, we compute the fractional contribution of \mychng{the flux in the continuum bands to the total FUV flux over the full range containing the continuum bands (\vcontWmin\ ~--~ \contWmax) and present the results in the last column of Table \ref{tbl:fuvcont}.
This fraction does not account for the continuum flux present within emission line regions.
Therefore, it serves as a lower limit on the total contribution of continuum within the \vcontWmin\ ~--~ \contWmax\ range.
This contribution is at least on the order of 10\%} for the stars where it is detected, and where it is not detected the uncertainties do not rule out similar levels.

\begin{deluxetable}{lrrr}
\tablewidth{0pt}
\tabletypesize{\footnotesize}
\tablecaption{FUV continuum measurements \label{tbl:fuvcont}}

\tablehead{
\colhead{Star} & \colhead{FUV Continuum} & \colhead{Detection} & \colhead{Fraction}\\
\colhead{} & \colhead{Flux} & \colhead{Significance} & \colhead{of Flux\tablenotemark{a}}\\
\colhead{} & \colhead{[erg s$^{{-1}}$ cm$^{{-2}}$ \AA$^{{-1}}$]} & \colhead{$\sigma$} & \colhead{}}
\startdata
GJ 1214 & $-1.2 \pm 1.1 \sn{-15}$ & \nodata & -0.753\\
GJ 876 & $6.7 \pm 1.2 \sn{-15}$ & 5.7 & 0.080\\
GJ 436 & $1.1 \pm 1.4 \sn{-15}$ & \nodata & 0.082\\
GJ 581 & $-8 \pm 13 \sn{-16}$ & \nodata & -0.201\\
GJ 667C & $1.1 \pm 1.1 \sn{-15}$ & \nodata & 0.088\\
GJ 176 & $4.1 \pm 1.2 \sn{-15}$ & 3.5 & 0.070\\
GJ 832 & $6.4 \pm 1.1 \sn{-15}$ & 6.1 & 0.122\\
HD 85512 & $6.4 \pm 5.4 \sn{-15}$ & \nodata & 0.114\\
HD 40307 & $8.2 \pm 4.8 \sn{-15}$ & \nodata & 0.147\\
HD 97658 & $1.13 \pm 0.18 \sn{-14}$ & 6.3 & 0.201\\
$\mathrm{\epsilon}$ Eri & $8.128 \pm 0.038 \sn{-13}$ & 215.9 & 0.128\\
\enddata

\tablenotetext{a}{Integral of flux within continuum bands divided by the integral of all flux over the range containing those bands.}

\end{deluxetable}

\subsection{Estimating SEDs for Stars not in the MUSCLES Catalog}
A means of estimating the UV SED for any star without \scraft{HST} observations would be very useful. 
We investigated the potential for using photometry in the GALEX FUV and NUV bands to empirically predict fluxes in other UV bands.
However, the GALEX All-sky Survey (AIS; \citealt{bianchi11}) contains magnitudes for only half of the MUSCLES sample. 
The MUSCLES stars represent the brightest known M and K dwarf hosts.
Therefore, they will be better represented than a volume-limited sample of such stars in the GALEX AIS catalog. 
That half do not appear in the catalog suggests the GALEX AIS catalog, while an excellent tool for population studies with few restrictions on sample selection, is not a generally useful tool for studies involving one or a few preselected targets.
Nonetheless, we checked for a correlation of GALEX FUV magnitudes for the 6 targets in the AIS catalog with their integrated FUV flux from \scraft{HST} data and found no correlation.
It bears mentioning that while most M and K dwarf planet hosts may not appear in GALEX catalogs, the GALEX survey data have proved useful both for prospecting for low-mass stars \citep{shkolnik11} and population studies of these stars \citep{shkolnik14,jones16}. 

There are other means, besides GALEX data, of estimating broadband and line fluxes for a low-mass star without complete knowledge of its SED. 
To this end, Paper I provides useful relations between broadband fluxes and emission line fluxes.
Paper I also quantifies levels of FUV activity, taken as the ratio of the integrated FUV flux to the bolometric flux, and shows that they are roughly constant.
This consistency suggests that the median ratio of FUV to bolometric flux  of the MUSCLES sample would be appropriate as a first-order estimate of that same ratio for any similar low-mass star.
Further, the MUSCLES spectra, normalized by their bolometric flux, can serve as rough template spectra for a low-mass star with no available spectral observations. 
If the star's bolometric flux is known, this can be multiplied with the template normalized spectrum to provide absolute flux estimates.

\refchng{21}{However, age is likely a factor in the consistency in FUV activity among MUSCLES targets.
Age estimates are not presently available for the full MUSCLES sample; however, ages of the MUSCLES stars are likely to be consistent with other nearby field stars of similarly weak \ion{Ca}{2} H and K flux (see Paper I).
These have ages of several Gyr \citep{mamajek08}.
Stars with ages under a Gyr are likely to have higher levels of FUV and NUV activity compared to the MUSCLES sample.
This conclusion follows from the recent work of \cite{shkolnik14} that demonstrated a dependence of excess (i.e. non-photospheric) FUV and NUV flux on stellar age for early M dwarfs. 
\cite{shkolnik14} found that the excess FUV and NUV flux of their stellar samples remained roughly constant at a saturated level until an age of a few hundred Myr.
After this age, the excess flux declines by over an order of magnitude as the stars reach ages of several Gyr.
Because of this age dependence in FUV flux, the consistency in FUV activity of the MUSCLES stars should not be taken to be representative of stars with ages under a Gyr.}

\subsection{Photodissociation}
\label{sec:dissrates}

The composition of a planetary atmosphere depends on a complex array of factors including mass transport, geologic sources and sinks, biological activity, impacts, aqueous chemistry, stellar wind, and incident radiation (e.g. \citealt{matsui86,lammer07,seager10,hu12,kaltenegger13}).
Many of these factors require assumptions weakly constrained by what is known of solar system planets. 
However, the MUSCLES data directly characterize the radiation field incident upon planets around the 11 host stars.  
This allows the use of MUSCLES data to address two top-level questions regarding atmospheric photochemistry without detailed modeling of specific atmospheres: (1) What is the relative importance of various spectral features to the photodissociation of important molecules, and (2) how does this differ between stars?

To quantitatively answer these questions, we examined the photodissociation rates of various molecules as a function of wavelength resulting from the radiative input of differing stars. 
Because we leave atmospheric modeling to future work, we did not introduce any attenuation of the stellar SED by intervening material.
In other words, we assumed direct exposure of the molecules to the stellar flux.
However, for wavelengths shortward of the ionization threshold of the molecule, we assumed that all photon absorptions result in ionization rather than dissociation.
Consequently, for each molecule we ignore stellar flux shortward of that molecule's ionization threshold.
Because we assumed direct exposure to the stellar flux, we refer to the photodissociation rates we have computed as ``unshielded.''

\subsubsection{Method of Computing ``Spectral'' Photodissociation Rates}
To determine ``spectral'' photodissociation rates, $\pds(\lambda)$, we multiplied the wavelength-dependent cross sections, $\xsctn(\lambda)$, by the photon flux density, $\pflux(\lambda)$, and the sum of the quantum yields for all dissociation pathways, $q(\lambda)$. The quantum yield of a pathway expresses the probability that the molecule will dissociate through that pathway if it absorbs a photon with wavelength $\lambda$. To wit,
\begin{equation}
\label{eq:pds}
\pds(\lambda) = \pflux(\lambda)\xsctn(\lambda)\sum_i{q_i(\lambda)}.
\end{equation}
Note that the stellar spectrum must be specified in \emph{photon} flux density [\pers\ cm$^{-2}$ \perAA] rather than energy flux density [\fluxcgs\ \perAA] ($\pflux = \eflux / (hc/\lambda)$). 

The photodissociation cross sections and quantum yields we used come from the data gathered by \cite{hu12} and presented in their Table 2, updated to include the high-resolution cross section measurements of \OII\ in the Schumann-Runge bands from \cite{yoshino92}.
We extrapolated the \HIIO\ cross section data, as is conventional, from $\sim$2000 \AA\ to the dissociation limit at 2400 \AA\ using a power-law fit to the final two available data points.
The cross sections only apply to photodissociation from the ground state.
We also do not include \HII\ dissociation by excitation of the Lyman and Werner bands. 
Physically, $\pds$ in Eq. \ref{eq:pds} represents the rate at which unshielded molecules are photodissociated by stellar photons per spectral element at $\lambda$. Thus the units of $\pds$ are \pers \perAA.

The absolute level of $\pds$ is tied to a specific distance from the star since $\pflux$ drops with the inverse square of distance.
For the values we present in the remainder of Section \ref{sec:dissrates}, we set the $\pflux$ level of each star's spectrum such that the bolometric \emph{energy} flux (``instellation''), $\ebolo = \int_0^\infty \eflux(\lambda)d\lambda$, was equivalent to Earth's insolation.

The dissociation rate, $\drate$ [\pers], due to radiation in the wavelength range $(\lambda_a, \lambda_b)$ is then given by 
\begin{equation}
\drate = \int_{\lambda_a}^{\lambda_b} \pflux(\lambda)\xsctn(\lambda)\sum_i{q_i(\lambda)}d\lambda.
\end{equation}
Integrating from the molecule's ionization threshold wavelength, $\lamion$, to $\infty$ gives the total dissociation rate due to irradiation from stellar photons (neglecting any dissociations from photons with sufficient energy to ionize the molecule).

\begin{deluxetable*}{lrrrrrrrrrr}
\centering
\tabletypesize{\scriptsize}
\tablewidth{0pt}
\tablecaption{Dissociation rates [\pers] for unshielded molecules receiving bolometric flux equivalent to Earth's insolation. \label{tbl:dissrates}}

\tablehead{\colhead{Star} & \colhead{H$_2$} & \colhead{N$_2$} & \colhead{O$_2$} & \colhead{O$_3$} & \colhead{H$_2$O} & \colhead{CO} & \colhead{CO$_2$} & \colhead{CH$_4$} & \colhead{N$_2$O} & \colhead{O$_2$/O$_3$\tnm{a}}\\\colhead{} & \colhead{$\sn{-7}$} & \colhead{$\sn{-6}$} & \colhead{$\sn{-6}$} & \colhead{$\sn{-4}$} & \colhead{$\sn{-5}$} & \colhead{$\sn{-6}$} & \colhead{$\sn{-5}$} & \colhead{$\sn{-5}$} & \colhead{$\sn{-7}$} & \colhead{}}
\startdata
GJ 1214 & 0.36 & 3.3 & 0.86 & 1.2 & 3.2 & 2.6 & 0.7 & 4.2 & 0.82 & 0.0069\\
GJ 876 & 0.49 & 2.6 & 4.2 & 1.4 & 3.0 & 2.0 & 0.54 & 4.1 & 2.4 & 0.031\\
GJ 436 & 1.2 & 3.7 & 1.2 & 1.9 & 3.3 & 2.6 & 0.69 & 4.5 & 0.85 & 0.0065\\
GJ 581 & 0.18 & 1.4 & 0.59 & 1.5 & 1.6 & 1.1 & 0.3 & 2.2 & 0.34 & 0.0039\\
GJ 667C & 3.1 & 8.9 & 1.6 & 3.0 & 7.6 & 6.2 & 1.6 & 10 & 1.1 & 0.0052\\
GJ 176 & 2.1 & 5.1 & 3.4 & 2.6 & 4.2 & 3.4 & 0.89 & 5.6 & 2.2 & 0.013\\
GJ 832 & 2.2 & 6.5 & 2.1 & 3.7 & 5.7 & 4.6 & 1.2 & 7.8 & 1.6 & 0.0057\\
HD 85512 & 6.6 & 8.1 & 1.1 & 5.6 & 3.9 & 4.5 & 1.0 & 5.4 & 1.1 & 0.002\\
HD 40307 & 17 & 17 & 1.1 & 15 & 5.5 & 8.0 & 1.6 & 7.5 & 1.4 & 0.00075\\
HD 97658 & 17 & 15 & 1.8 & 34 & 4.9 & 7.2 & 1.4 & 6.6 & 3.6 & 0.00052\\
$\mathrm{\epsilon}$ Eri & 40 & 33 & 5.0 & 29 & 8.4 & 15 & 2.7 & 11 & 5.4 & 0.0017\\
Sun & 0.56 & 1.1 & 2.5 & 81 & 1.1 & 0.67 & 0.13 & 0.88 & 13 & 0.0003\\
\enddata

\tablenotetext{b}{Ratio of the \OIII\ to \OII\ dissociation rates.}

\end{deluxetable*}

To examine the importance of various spectral regions and features to dissociating a given molecule, a useful quantity is the ``cumulative distribution'' of $j$, normalized by the total dissociation rate, i.e.
\begin{equation}
\cds = \frac{\int_{\lamion}^{\lambda} \pds(\lambda')d\lambda'}{\int_{\lamion}^{\infty} \pds(\lambda')d\lambda'}.
\end{equation}
where $\lambda_\mathrm{ion}$ is the ionization threshold of the molecule.
Physically, $\cds$ expresses the fraction of the total photodissociation rate of a molecule that is due to photons with wavelengths between the molecule's ionization threshold and $\lambda$.
This permits easy visual interpretation from plotted $\cds$ values of the degree to which a portion of the stellar spectrum is driving photodissociation of a molecule. 
The change in $\cds$ over a given wavelength range gives the fraction of all dissociations attributable to photons in that range.
Thus, ranges where $\cds$ rapidly rises by a large amount indicate that stellar radiation with wavelengths in that range contribute disproportionately to the overall rate of photodissociation.

\subsubsection{Comparison of Unshielded Photodissociation Rates of \moleculelist\ Among the MUSCLES Stars}

For all 11 MUSCLES hosts and the Sun, we have tabulated the total unshielded dissociation rate, $\dratebol$, of the molecules \moleculelist\ in Table \ref{tbl:dissrates}.
We considered these molecules because of their ubiquity and biological relevance.
Among the MUSCLES stars, unshielded photodissociation rates of most of these molecules range over roughly an order of magnitude.
Exceptions are \HII, for which $\dratebol$ values vary by over two orders of magnitude, and \HIIO\ and \CHIV, for which $\dratebol$ values vary by about a factor of 5. 
Confining the comparison to only the 4 K stars or only the 7 M stars reduces the spread in $\dratebol$ values, but the values still vary by a factor of a few or more. 
This variability between stars emphasizes the importance of spectral observations of individual exoplanet host stars to photochemical modeling.

Median unshielded photodissociation rates for the MUSCLES stars are generally a factor of a few higher than the Sun. 
However, one key exception to this is \OIII, for which the Sun's relatively high level of photospheric flux in the NUV makes it the strongest dissociator by a factor of a few to nearly two orders of magnitude. 
We reserve further discussion of this molecule for Section \ref{sec:o3}.

\begin{figure*}
\centering
\includegraphics{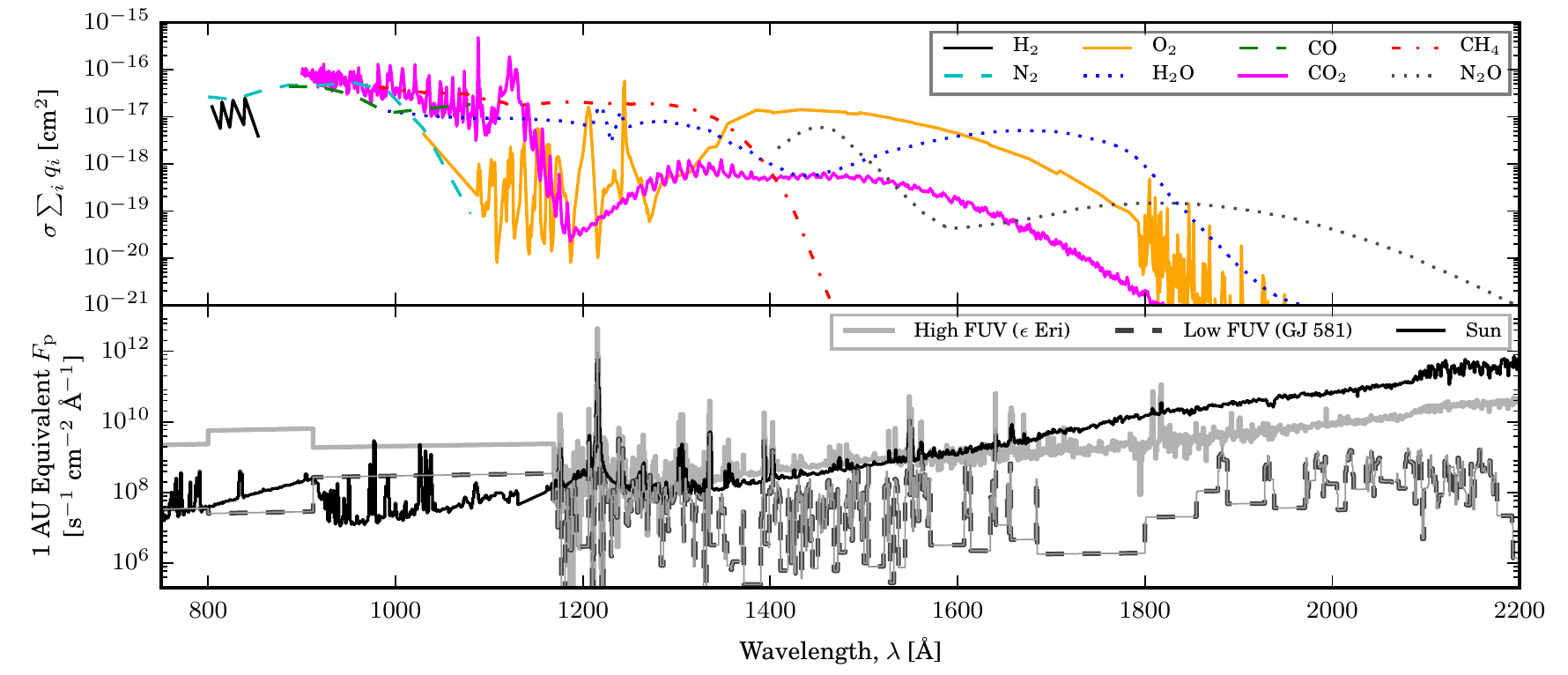}
\caption{Top: Photodissociation cross sections of the examined molecules.
Bottom: SEDs of the reference stars: the most active MUSCLES star (as defined by the ratio of FUV to bolometric flux), least active MUSCLES star, and the Sun, converted to photon flux density instead of energy flux density and scaled such that the bolometric energy flux is equivalent to Earth's insolation. }
\label{fig:xspecs}
\end{figure*}

\begin{figure}
\centering
\includegraphics{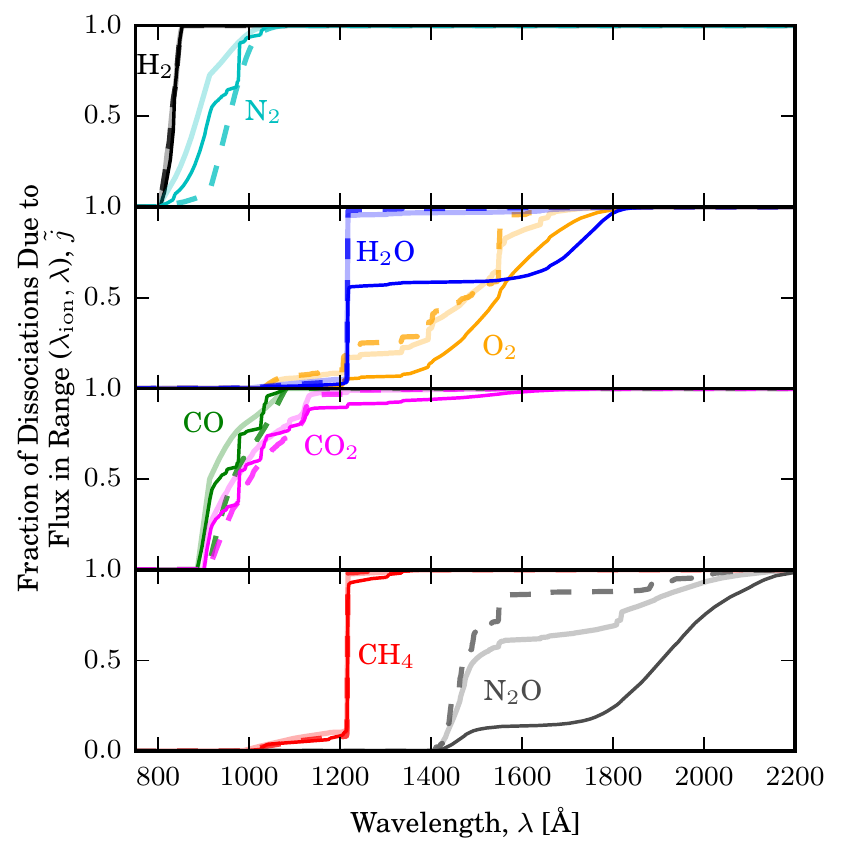}
\caption{Cumulative photodissociation spectra.
The curves show what fraction of the dissociation is due to photons with wavelengths from the molecule's ionization threshold ($\lamion$) to $\lambda$.
This corresponds to the product of the curves in Figure \ref{fig:xspecs} integrated from $\lamion$ to $\lambda$ and normalized by the full integral.
The curve's rate of growth indicates the importance of that region of the spectrum to photodissociation of the molecule.
Each molecule has three curves corresponding to the three reference stars whose spectra are plotted in the bottom panel of Figure \ref{fig:xspecs} using the same line styles. 
Groups of curves are labeled by molecule and colored to match the top panel of Figure \ref{fig:xspecs}.}
\label{fig:dissrates}
\end{figure}

To examine the wavelength dependence of unshielded photodissociation rates, we computed $\cds$ curves for three representative stars: the most active star in the MUSCLES sample, \epseri; the least active star, GJ 581; and the Sun, where we define activity as the ratio of the integrated FUV flux to the bolometric flux.
Before presenting plots of the computed $\cds$ curves, we first plot in Figure \ref{fig:xspecs} the photodissociation cross sections incorporating quantum yields, $\xsctn\sum_i{q_i(\lambda)}$, and the stellar photon flux density, $\pflux(\lambda)$.
The spectrum of GJ 581 had very low S/N in some areas, resulting in many spectral bins with negative flux density values due to the subtraction of background count rate estimates. 
Thus, for all stars we merged any negative-flux bins with their neighbors until the summed flux was above zero.
This amounts to sacrificing resolution for S/N in these areas.
The integrated and normalized product of the two curves in Figure \ref{fig:xspecs} then gives the $\cds$ curves plotted in Figure \ref{fig:dissrates}. 
Because $\cds$ values are normalized by the overall photodissociation rate, $\dratebol$, for a given input SED, Figure \ref{fig:dissrates} does not show how $\dratebol$ varies between stars.
\refchng{23}{However, this information is available in Table \ref{tbl:dissrates}.}

From Figure \ref{fig:dissrates} it is clear that, with the exception of \HIIO\ and \NIIO, the same portions of the stellar spectra drive unshielded photodissociations of each molecule.
This results from the relatively small differences in the reference star SED shapes over the range of photon wavelengths capable of photodissociating these molecules.

Although similar between stars for most molecules, the $\cds$ curves reveal some intriguing structure.
The dissociation of \HII, \NII, CO, and \COII\ is primarily due to flux shortward of the \cosCutLo\ cutoff of the COS data.
\refchng{25}{There is an apparent difference between the curves for the Sun versus GJ 581 and \epseri\ in this region:
The curve for the Sun shows jumps not present in the curves of GJ 581 and \epseri.
The jumps are caused by solar line emission that significantly dissociates these molecules.
The jumps are not present in the curves for GJ 581 and \epseri\ because these stars were not directly observed at those wavelengths.
Flux in this region is instead given by the broadband estimates described in Section \ref{sec:lyareduction} that lack any individual spectral lines.}
\mychng{However, it is probable that line emission from GJ 581 and \epseri\ in these ranges will}, like the Sun, be important to the photodissociation of \HII, \NII, CO, and \COII.
Similarly, note that while the \COII\ $\cds$ curves of \epseri\ and GJ 581 show some jumps, these are due to spikes in the photodissociation cross section of \COII\ rather than structure in the \epseri\ and GJ 581 SEDs.

For \OII, the photodissociation cross section has a narrow peak near \lya\ and a broad, highly asymmetric peak at $\sim$1400 \AA. 
Thus, for the low-mass stars, emission from the \lya; \Siiv\ $\lambda\lambda$ 1393,1402; and \Civ\ $\lambda\lambda$ 1548,1550 lines contributes significantly to the overall photodissociation of \OII.
These are weaker relative to the FUV continuum in the Sun, so instead the solar FUV continuum dominates photodissociation.

Unshielded photodissociation of \HIIO\ is dominated by \lya\ in \epseri\ and GJ 581 and unshielded photodissociation of \CHIV\ is dominated by \lya\ for all three reference stars. 
For the Sun, dissociation of \HIIO\ is roughly evenly shared between \lya\ and the 1600~--~1800 \AA\ FUV because of weaker \lya\ emission and stronger FUV continuum emission. 
However, when interpreting the effect of \lya\ emission on photodissociation rates, it is especially important to consider the assumption of unshielded molecules.
\lya\ is significantly attenuated by only small amounts of intervening \Hi, \HII, \HIIO, or \CHIV.
These species scatter and absorb \lya\ so readily that their presence limits \lya\ dissociation to the uppermost reaches of an atmosphere.
For reference, unity optical depth for the center of the \lya\ line occurs at a column density of $3\sn{17}$ cm$^{-2}$ of \HII\ from scattering and column densities of $7\sn{16}$ cm$^{-2}$ of \HIIO\ or $5\sn{16}$ cm$^{-2}$ of \CHIV\ from absorption.
These densities correspond to pressures of order $10^{-9}$ bar for planets with surface gravities close to Earth's. 
For an investigation of the effect of \lya\ on mini-Neptune atmospheres, see \cite{miguel15}.

Like \HIIO, \NIIO\ dissociation is driven by differing wavelength ranges when comparing the SED of the Sun with that of \epseri\ and GJ 581.
While the photodissociation cross section of \NIIO\ peaks in the FUV at around 1450 \AA, a secondary peak that is several times broader and some two orders of magnitude lower occurs in the NUV near 1800 \AA.
For \epseri\ and GJ 581, flux levels are slow to rise across the region encompassing these peaks, so radiation at the 1450 \AA\ primary peak dominates dissociations.
In contrast, the solar spectrum rises very rapidly over the same range and beyond.
Consequently, solar radiation from $\sim$1800-2100 \AA\ dominates the dissociation of \NIIO.

Overlying material will attenuate the stellar flux density at all wavelengths within an atmosphere, but the degree of attenuation varies by many orders of magnitude across the spectrum.
This modifies both the overall level and the spectral content of the ambient radiation field with atmospheric depth.
In turn, the shapes of the $\cds$ curves in Figure \ref{fig:dissrates} change and values of $\dratebol$ diminish as atmospheric depth is increased.
Many atmospheric models incorporate a treatment of radiative transfer in order to address the change in the ambient radiation field with atmospheric depth and accurately incorporate photochemistry (see, e.g., \citealt{segura05, hu12, grenfell14, rugheimer15}).
Although inapplicable within an atmosphere, the unshielded photodissociation rates we present allow a physical comparison of ``top of the atmosphere'' photochemical forcing as a function of stellar host and wavelength.

\subsubsection{Abiotic \OII\ and \OIII\ Production and the Significance of Visible Radiation in \OIII\ Photodissociation}
\label{sec:o3}

The photodissociation of \OII\ and \OIII\ is of special interest because of the potential use of these molecules as biomarkers (e.g. \citealt{lovelock65,selsis02,tian14,harman15}).
Significant abiotic production of these molecules is possible in \COII -rich atmospheres through \COII\ dissociation that liberates free O atoms to combine with O and \OII, creating \OII\ and \OIII\ \citep{selsis02,hu12,tian14,domagal14,harman15}.
Abiotic \OII\ and \OIII\ generation is also possible in water dominated atmospheres  from \HIIO\ photolysis and subsequent H loss \citep{wordsworth14}.
Further, \OII\ and \OIII\ buildup by water photolysis is especially important early in the life of a low-mass star when XUV and UV fluxes are higher \citep{luger15b}.
\cite{hu12} found that the strongest factor controlling abiotic \OII\ levels is the presence of reducing species (such as outgassed \HII\ and \CHIV), extending the previous work by \cite{segura07} to planetary scenarios with very low volcanic outgassing rates.
In addition \cite{tian14} and \cite{harman15} found that the ratio of FUV to NUV flux positively correlates with abiotic \OII\ and \OIII\ abundances.
This ratio is tabulated for the MUSCLES stars in Paper I. 
\cite{tian14} and \cite{harman15} suggest that the dependence of abiotic \OII\ abundance on the FUV/NUV ratio results from a variety indirect pathways according to the various atmospheric compositions considered.

In contrast to the indirect effect the FUV/NUV ratio has on \OII\ levels, it has a direct impact on the production of \OIII\ in \COII -rich or \OII -rich atmospheres.
NUV photons dissociate \OIII, whereas FUV photons liberate free O atoms from \OII\ and \COII\ that can then combine in a three-body reaction with \OII\ to create \OIII.

Photodissociation rates vary among MUSCLES stars by over an order of magnitude for \OII\ and \OIII\ (Table \ref{tbl:dissrates}).
As a proxy for the relative forcing to higher \OIII\ populations in an \OII\ -rich atmosphere, we give the ratio of the unshielded dissociation rate of \OII\ to \OIII\ in Table \ref{tbl:dissrates}.
Higher ratios indicate photochemical forcing towards larger \OIII\ concentrations through more rapid dissociation of \OII\ or less rapid dissociation of \OIII.
These ratios vary by factors of a few in the MUSCLES stars, with \epseri\ standing out as having a ratio roughly an order of magnitude below the median.
The Sun has comparatively weak \OIII\ forcing -- two orders of magnitude below the median.

\begin{figure}
\centering
\includegraphics{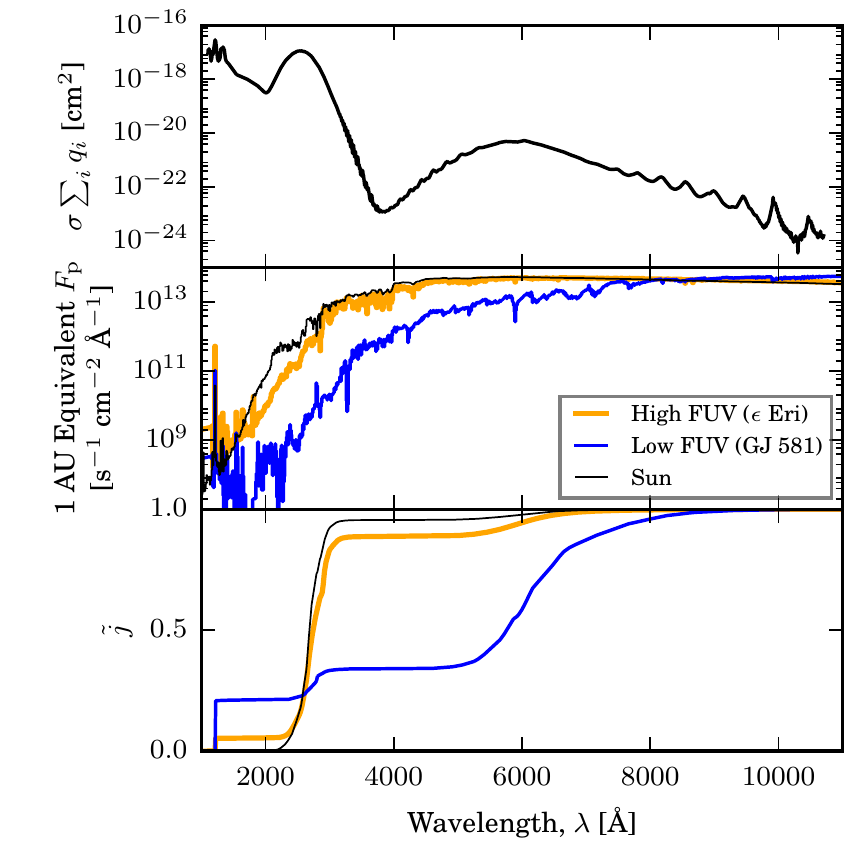}
\caption{Same as Figures \ref{fig:xspecs} and \ref{fig:dissrates} for \OIII. 
Top: Photodissociation cross section of \OIII. 
Middle: Photon flux density spectra of \epseri, GJ 581, and the Sun scaled such that the bolometric energy flux is equivalent to Earth's insolation. 
Bottom: Fraction of \OIII\ dissociations due to photons with wavelengths between $\lamion$ and $\lambda$ if exposed to the spectra in the middle panel.}
\label{fig:o3specs}
\end{figure}

Interestingly, NUV flux does not dominate the dissociation of \OIII\ in all stars.
For those with weak levels of NUV flux, visible light can play a roughly equal role.
This is apparent when examining photodissociation of \OIII\ by the unattenuated radiation of \epseri\ and GJ 581.
Figure \ref{fig:o3specs} shows this, plotting the photodissociation cross section of \OIII, photon flux density of the reference stars, and the resulting $\cds$ curves for \OIII\ (akin to Figures \ref{fig:xspecs} and \ref{fig:dissrates}).
For \epseri\ and the Sun, 2500~--~3200 \AA\ NUV flux is responsible for the bulk of \OIII\ dissociation. 
In stark contrast, for GJ 581 about 20\% of the dissociation is a result of \lya, and visible photons are responsible for most of the remainder.
The NUV flux of GJ 581, at roughly two orders of magnitude below that of the Sun and \epseri, provides a minimal contribution, about 10\%, to the photolysis of \OIII.

\begin{figure}
\centering
\includegraphics{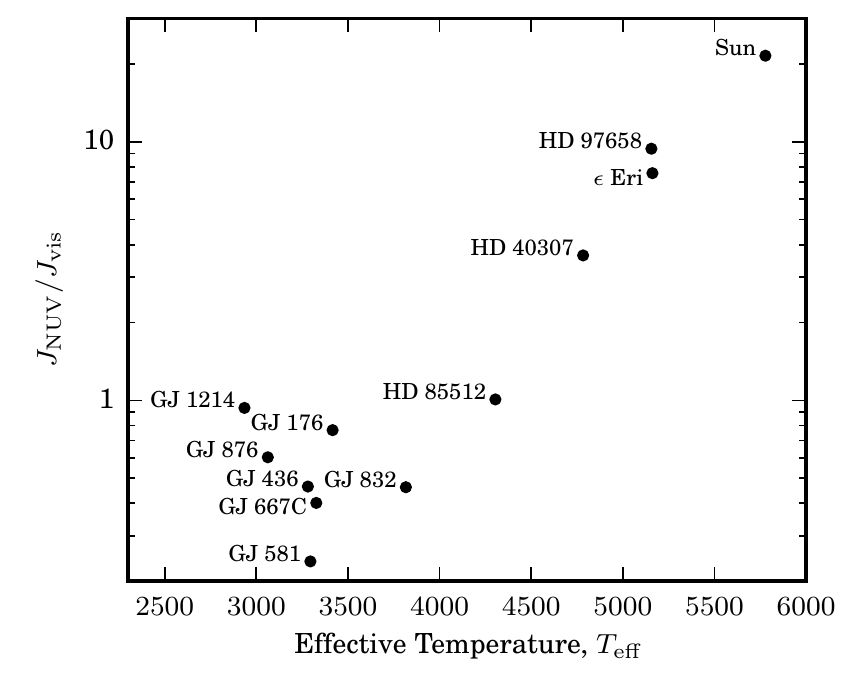}
\caption{Ratio of dissociations of \OIII\ from NUV photons to those from visible photons if exposed to the unattenuated flux of the MUSCLES stars versus \mychng{the stellar effective temperatures we estimated.} 
Visible flux is important for all stars with \teff\ under 4500 K, but this importance diminishes with increasing \teff\ above 4500 K.}
\label{fig:o3dissratios}
\end{figure}

\begin{figure*}
\centering
\includegraphics{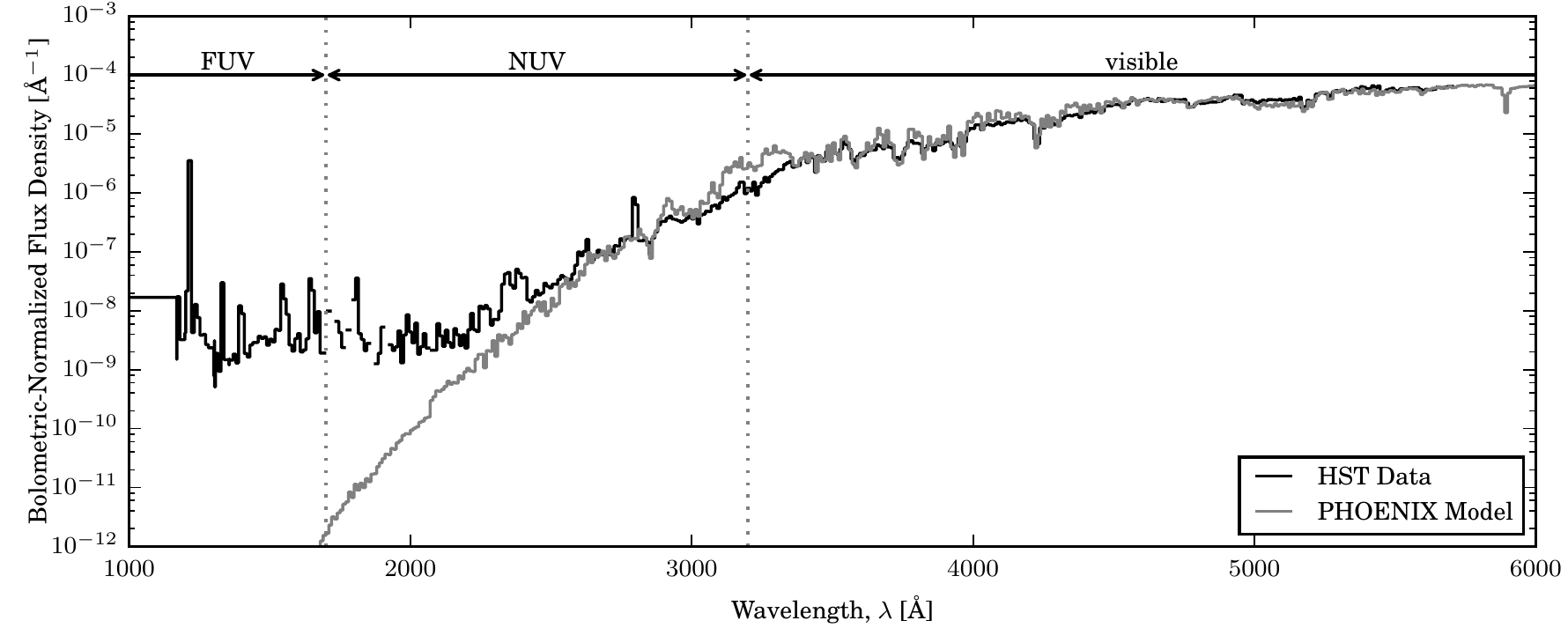}
\caption{Fit of the PHOENIX model spectrum for \phxCompareStar. 
The spectra are identically binned to facilitate comparison.
The PHOENIX models closely track the visible spectrum measured by \scraft{HST}, but poorly predict the flux in the FUV and NUV. 
The discrepancy near 3200 \AA\ dominates the integrated NUV flux, causing a factor $\sim 2$ overprediction of the NUV by the PHOENIX models. 
The PHOENIX FUV flux is negligible compared to that measured by \scraft{HST}.}
\label{fig:phx_comparison}
\end{figure*}

Indeed, visible radiation plays an important role in unshielded photodissociation of \OIII\ for all of the MUSCLES M dwarfs.
This pattern is depicted in Figure \ref{fig:o3dissratios}, comparing the ratio of \OIII\ photolysis from NUV flux to visible flux with the stellar effective temperature. 
For the M dwarfs, photospheric continuum flux is no longer a significant contributor in the NUV and visible radiation contributes as much or more than NUV flux to the photolysis of \OIII.
\mychng{This suggests that, among M dwarf hosts, the ratio of FUV to visible flux might prove a better predictor of relative \OIII\ populations in planetary atmospheres than the FUV/NUV ratio.}

\subsection{A Note on Models of Stellar Atmospheres}
\label{sec:badphoenix}
Section \ref{sec:dissrates} demonstrates the importance of including accurate levels of the stellar FUV flux in the SEDs used as input to photochemical models.
When such models involve a system lacking observations of the host star, the modeler must draw from other sources to provide the necessary input SED. 
One such source is the several existing codes that can synthesize a stellar SED by modeling the stellar atmosphere, most notably the PHOENIX code \citep{hauschildt99}.
However, most implementations of these codes truncate the stellar atmosphere above the photosphere. This includes the \cite{husser13} code that generated the spectra covering the visible and IR portions of the MUSCLES SEDs.
Such photosphere-only models greatly underestimate levels of FUV flux emitted by low-mass, cool stars.

The extremity of this underestimate is illustrated in Figure \ref{fig:phx_comparison} by plotting the panchromatic SED of \phxCompareStar, incorporating the UV observations by \scraft{HST}, against the photosphere-only PHOENIX spectrum interpolated from the \cite{husser13} catalog. 
Comparing the two spectra shows that the \cite{husser13} PHOENIX spectrum  predicts a flux lower than the observed value by three orders of magnitude or greater throughout the FUV. 
A similar discrepancy has been noted before for stars of mass equal to or less than that of the Sun \citep{franchini98, seager13, shkolnik14, rugheimer15}.
Because the FUV is where the photodissociation cross sections of many molecules peak, the use of inaccurately low FUV flux levels could underpredict rates of photodissociations if, like the MUSCLES stars, most low-mass stars show levels of UV emission exceeding that predicted by photosphere-only models. 
Similarly, photosphere-only models are likely to underpredict EUV levels and associated heating rates in planetary thermospheres by many orders of magnitude.

The prediction of lower than observed levels of short-wavelength flux is not a oversight of the \cite{husser13} PHOENIX model. 
Rather, the \cite{husser13} models and most others omit the stellar upper atmosphere by design. 
While not common, models that attempt to reproduce emission from the stellar upper atmosphere do exist.
The same PHOENIX code often used for modeling photospheric emission has also been adapted for chromospheric emission from five M dwarfs \citep{fuhrmeister05} and for CN Leonis during a flare \citep{fuhrmeister10}. 
In addition, much work on chromospheric modeling has been done by the Houdebine group, beginning with \cite{houdebine94}. 
These works focused on ground-observable NUV and visible emission, rather than the FUV emission most important to photochemistry. 
Alternatively, a model by \citeauthor{fontenla16} (\citeyear{fontenla16}; submitted) reproduces much of the spectral structure of the X-ray, UV, and visible emission of the M1.5V star GJ 832, and work underway by \cite{peacock15} seeks to model the same emission for a variety of M dwarf stars. 
Generalized to a broader range of stellar types, we expect models like this will be indispensable for the simulation of photochemistry in the atmospheres of planets orbiting stars lacking detailed observations.
These models require a sample of stars with detailed UV observations for validation. 
The MUSCLES Treasury Survey provides this sample.

\section{Summary}
\label{sec:summary}

We have presented a catalog of SEDs spanning X-ray to IR wavelengths for 11 nearby, low-mass exoplanet host stars. 
Using these spectra, we examined the line versus continuum emission energy budget in the FUV, detecting an inter-line continuum at $>3\sigma$ confidence for \nContDetections\ stars. 
This revealed a likely ionization edge structure in the FUV continuum of \epseri\ that could represent the \Siii\ recombination continuum. 
\mychng{We found that, when it was detected, the continuum contributed around 10\% of the flux in the range spanned by the continuum bands.}
Uncertainties on nondetections were also consistent with this level.

Additionally, we examined photodissociation rates of the molecules \moleculelist\ assuming that these species were exposed directly to the unattenuated stellar flux. 
We found total dissociation rates driven by different stars in our sample varied by over an order of magnitude for the majority of these molecules.
By comparison, the photodissociation rates for the solar spectrum are near or below the lowest values for MUSCLES stars, except for \OIII\ where the solar photospheric NUV flux produces dissociation rates several times larger than the median value for MUSCLES stars. 
We also examined the relative importance of different spectral regions to the photodissociation of unshielded molecules driven by the SEDs of the most active host star (\epseri), least active host star (GJ 581), and the Sun.
There is little variation among these stars in which portion of the stellar emission is responsible for the bulk of unshielded photodissociations for most of the molecules examined.
However, for \OIII\ we found that the dominant dissociative wavelength range can move from the NUV to the visible for low-mass stars with little NUV flux.

The spectral catalog presented in this paper, \href{\hlsplink}{archived with MAST}\footnote{\hlsplink}, provides critical data for vetting stellar models and simulating photochemistry in planetary atmospheres. 

\section{Acknowledgments}
The data presented here were obtained as part of the HST Guest Observing programs \#12464 and \#13650 as well as the COS Science Team Guaranteed Time programs \#12034 and \#12035. 
This work was supported by NASA grants HST-GO-12464.01 and HST-GO- 13650.01 to the University of Colorado at Boulder. 
We thank Tom Woods and Chris Moore for useful discussions that provided a solar context to the work.
This publication makes use of data products from the Two Micron All Sky Survey, which is a joint project of the University of Massachusetts and the Infrared Processing and Analysis Center/California Institute of Technology, funded by the National Aeronautics and Space Administration and the National Science Foundation.
Sarah Rugheimer would like to acknowledge support from the Simons Foundation (339489, Rugheimer). FT is supported by the National Natural Science Foundation of China (41175039) and the Startup Fund of the Ministry of Education of China (20131029170). This work was partially supported by Chandra grants GO4-15014X and GO5-16155X from Smithsonian
Astrophysical Observatory and NASA XMM grant NNX16AC09G.

\bibliography{refs_ADS,refs_other,refs_photometry}{}
\bibliographystyle{aasjournal}

\end{document}